\documentclass{aa}  

\usepackage{graphicx}
\usepackage{txfonts}
\usepackage{newtxtext,newtxmath}
\usepackage[autostyle]{csquotes}
\usepackage{subfig}
\usepackage{xcolor}
\usepackage{hyperref}
\usepackage{comment} % to be removed
\usepackage{multirow}
\usepackage{upgreek}
\begin{document}

   \title{Revealing the intricacies of radio galaxies and filaments in the merging galaxy cluster Abell 2255}
   \subtitle{II. Properties of filaments using multi-frequency radio data}
   
   \titlerunning{High-resolution radio filaments in Abell 2255}
   
   \authorrunning{De Rubeis et al.}

   \author{E. De Rubeis\inst{1,2}\thanks{Corresponding author: emanuele.de.rubeis@uni-hamburg.de},
            M. Bondi\inst{2},
            A. Botteon\inst{2},
            R. J. van Weeren\inst{3},
            J. M. G. H. J. de Jong\inst{3,4},
            G. Brunetti\inst{2},
            L. Rudnick\inst{5},
            M. Br\"uggen\inst{6},
            L. Bruno\inst{2},
            E. L. Escott\inst{7},
            C. Gheller\inst{2},
            L. K. Morabito\inst{7,8},
            K. Rajpurohit\inst{9},
            H. J. A. R\"ottgering\inst{3}}

   \institute{Dipartimento di Fisica e Astronomia, Universit\`a di Bologna, via Gobetti 93/2, I-40129 Bologna, Italy
    \and
    INAF - Istituto di Radioastronomia di Bologna, Via Gobetti 101, 40129 Bologna, Italy
    \and
    Leiden Observatory, Leiden University, PO Box 9513, NL-2300 RA Leiden, The Netherlands
    \and
    ASTRON, The Netherlands Institute for Radio Astronomy, Postbus 2, 7990 AA Dwingeloo, The Netherlands
    \and
    Minnesota Institute for Astrophysics, University of Minnesota, 116 Church St. SE, Minneapolis, MN 55455, USA
    \and
    Hamburger Sternwarte, Universit\"at Hamburg, Gojenbergsweg 112, 21029 Hamburg, Germany
    \and
    Centre for Extragalactic Astronomy, Department of Physics, Durham University, South Road, Durham DH1 3LE, UK
    \and
    Institute for Computational Cosmology, Department of Physics, Durham University, South Road, Durham DH1 3LE, UK
    \and
    Center for Astrophysics | Harvard \& Smithsonian, 60 Garden St., Cambridge, MA 02138, USA}

   \date{Received XXX; accepted YYY}

% \abstract{}{}{}{}{} 
% 5 {} token are mandatory
 
  \abstract
  % context heading (optional)
  % {} leave it empty if necessary
   {Thin, elongated, non-thermal filaments in galaxy clusters and groups are nowadays ubiquitous in sensitive radio maps. Despite the large (and increasing) number of cases, their origin is still unclear. In a previous work, we revealed a wealth of filaments surrounding the main member radio galaxy of Abell 2255: a head-tail named \enquote{Original Tailed Radio Galaxy} (Original TRG). We did this using 56 hours of sub-arcsecond resolution LOFAR-VLBI observations at 144~MHz.}
  % aims heading (mandatory)
   {In this paper, we aim to further analyze the filaments in Abell 2255 combining LOFAR data with uGMRT (1260~MHz) and VLA (1520~MHz) data to constrain the spectral shape of the filaments. This allows studying their morphological properties, required to understand their origin, at unprecedentedly high resolution ($\sim2.3$ kpc), crucial to disentangle the different cosmic ray components that populate the Original TRG.}
  % methods heading (mandatory)
   {We produced a LOFAR-VLBI map at $1.5^{\prime\prime}$ resolution using the wide-field technique with 56 hours of observations. This was the first time this technique was used for a galaxy cluster, especially for such deep observations. uGMRT and VLA data have been calibrated and imaged to produce spectral index maps and to apply further techniques to extract additional information, such as the radiative ages of the filaments or their equipartition magnetic field. Polarization information was also obtained using VLA through the rotation measure synthesis technique.}
  % results heading (mandatory)
   {Thanks to the LOFAR-VLBI wide-field image at 144~MHz, we revealed additional, very steep ($\alpha > 2$) filaments beyond those attached to the radio galaxy, extending for around 250 kpc and previously known as the Trail. Combining LOFAR-VLBI with uGMRT and VLA, we found integrated spectral values between $1.1-1.7$ for the filaments. Spectral analysis revealed also that the Original TRG has a complex structure, showing overlapping features with distinct spectral indices that extend throughout its tail. Polarized emission emerges only from the tail and the brightest part of the filaments, with values up to $22\%$. Although there is no clear scenario regarding the formation of filaments, we highlight the importance of the Original TRG as the main driver of such structures, even at larger distances from the core.}
  % conclusions heading (optional), leave it empty if necessary 
   {}


   \maketitle
%
%-------------------------------------------------------------------

\section{Introduction}
\label{sec:introduction}
Abell 2255 (hereafter A2255) is a nearby~\citep[$z = 0.0806$,][]{struble1999} merging galaxy cluster. In the past, it has been widely observed at radio frequencies because of the complex variety of emission on multiple scales related, among others, to the cluster member radio galaxies~\citep{harris1980, burns1995, miller2003, govoni2006, pizzo2009, pizzo2011, ternidegregory2017, botteon2020, botteon2022}. Moving within the intracluster medium (ICM), the synchrotron emitting jets ejected from the host galaxy are bent by ram pressure effects, while buoyancy effects can cause them to move towards the edge of the cluster~\citep{gunn1972, 1980miley, 2020hardcastle}. Based on the observed bending angle of the two jets, the radio morphology varies from wide-angle tails (WAT) to narrow-angle tails (NAT) to head-tails (HT), where the jets are bent in a common direction following the wake of the host galaxy, with the latter representing the head. In~\citet{derubeis2025}, hereafter~\hyperref[derubeis2025]{Paper I}, we presented deep (56 hours), International LOw Frequency ARray (LOFAR) telescope~\citep[ILT,][]{vanhaarlem2013} observations at 144~MHz of the main tailed radio galaxies in A2255. Thanks to the high-sensitivity and sub-arcsecond resolution ($0.3^{\prime\prime}-0.5^{\prime\prime}$), we resolved a wealth of filamentary emission (labeled in Fig.~\ref{fig:lofar_widefield_zoom} as in~\hyperref[derubeis2025]{Paper I}), especially surrounding the wake of the Original tailed radio galaxy (Original TRG), a NAT radio galaxy projected close to the cluster center. We characterized the morphological properties of these filaments, having lengths of $80-110$ kpc and varying widths between $3-10$ kpc. Given that their length is directly related to the driving scales of the turbulent flow~\citep{2018vazza}, we found that the dynamical lifetime of the filaments is comparable with the synchrotron radiative lifetime of electrons ($10^7-10^8$ yr), meaning that these can survive long enough to display radio emission before being dissipated by turbulence.
\par Such filaments are becoming increasingly frequent in radio images thanks to the combination of high-sensitivity and resolution of current interferometers, especially in the tails and in the surroundings of radio galaxies in cluster and group environments. Overall, they are characterized by projected lengths of $\rm{10s}-\rm{100s}$ kpc, widths from hundreds pc to a few kpc, and steep spectral indices ($\alpha > 1.3$~\footnote{We use the convention $S(\nu) \propto \nu^{-\alpha}$ for synchrotron spectrum, with $\alpha > 0$}). In some cases, significant spectral variations were observed along the filaments' length~\citep{rudnick2022,velovic2023}, although this is not ubiquitous~\citep[e.g.,][]{giacintucci2022, brienza2025, bushi2025}. These phenomena provide new opportunities for studying the physical processes that are occurring in the ICM, such as the magnetic field amplification or cosmic ray acceleration and propagation~\citep{bell2019}.
\par In this paper, we aim to study the physical properties of the filaments in A2255 combining LOFAR Very Long Baseline Interferometry (LOFAR-VLBI) image at 144~MHz with data from the upgraded Giant Meterwave Radio Telescope (uGMRT) at 1260~MHz and the Karl G. Jansky Very Large Array (VLA) at 1520~MHz. Data at higher frequencies are required to obtain spectral and polarization information on such structures, given that such structures have shown to be highly polarized in some cases~\citep[e.g.,][]{condon2021,rudnick2022,cotton2025}, while maintaining a sufficiently high angular resolution ($1.5^{\prime\prime}$) to disentangle all the various components already observed in Paper I. The emission from these filaments is expected to be strongly polarized due to the magnetic field being ordered along these structures. To improve the detection of the more extended filaments, linked to the upper end of the Original TRG and observable only at LOFAR frequencies~\citep[like the so-called Trail and T-bone,][]{botteon2020}, we used the LOFAR-VLBI wide-field technique, which enables imaging a field of view (FoV) up to $2.5^{\circ} \times 2.5^{\circ}$ using LOFAR international stations down to sub-arcsecond resolution with the high-band antenna (HBA). After its first employment by~\citet{sweijen2022}, to image the Lockman Hole field, this method was improved by recent works for the ELAIS-N1~\citep{ye2024,dejong2024,dejong2025calibration} and Bo\"otes (Escott et al., \textit{submitted}) fields. For A2255, we used an intermediate resolution of $1.5^{\prime\prime}$, required to match that of higher frequency observations, combining all the 56 hours of observations already used for Paper I, imaging a region of $1.5^{\circ} \times 1.5^{\circ}$ around the cluster. This resolution is optimal to recover the more extended emission (such as the one from the filaments, especially in the low signal-to-noise regime), and combined with smaller image size (with respect to previous works) allows to reduce the imaging computational time, which is the true bottleneck of the wide-field data processing~\citep[about 80\% of the total time for full resolution imaging, as found in][]{dejong2024}. This paper represents the first time this technique was used to target a galaxy cluster.
\par In this paper, we assume a flat $\rm{\Lambda}CDM$ cosmology, with $ H_{0} = 70~\rm{km~s^{-1}~Mpc^{-1}}$, $\Omega_{m} = 0.3$, and $\Omega_{\Lambda} = 0.7$. At the redshift of A2255, $1^{\prime\prime}$ corresponds to a linear scale of 1.512 kpc.

\section{Data calibration and imaging}
\label{sec:data_calibration}
\begin{figure*}[h!]
\centering
\includegraphics[width=\textwidth]{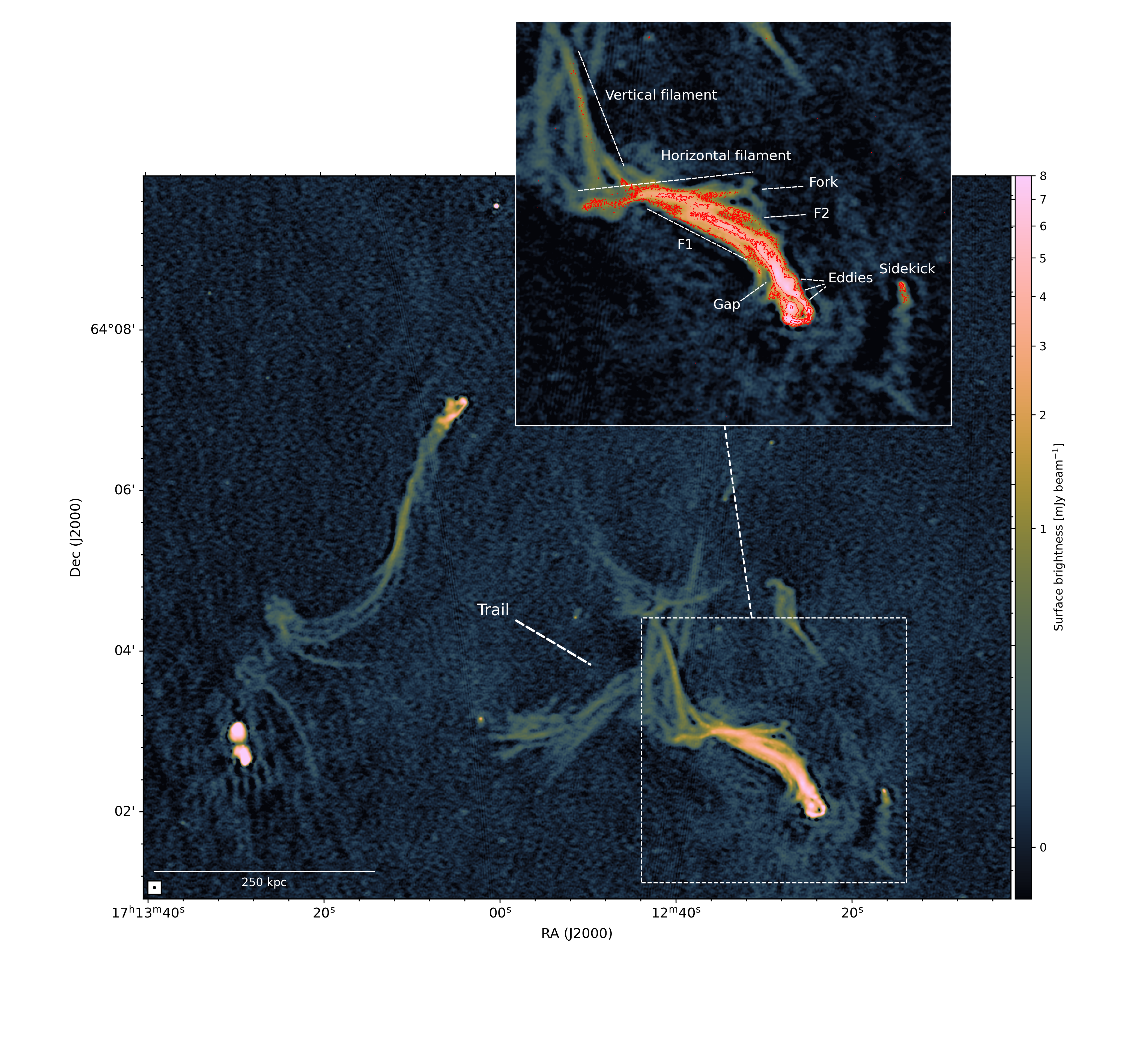}
\caption{LOFAR-VLBI wide-field image of the central region (size $9.2^\prime \times 7.5^\prime$) of A2255 at 144~MHz (from Fig~\ref{fig:lofar_widefield}). The final image has a resolution of $1.5^{\prime\prime}$, with rms noise of $38~\rm{\muup Jy~beam^{-1}}$. The restoring beam is shown in the bottom-left corner. Top panel: zoom on the Original TRG with the main features labeled as in Paper I. The contour levels from the sub-arcsecond resolution image from Paper I (Fig. 4 in that paper) are drawn in red at $3, 9, 27, 81, 243\sigma$ levels, with $\sigma=18~\rm{\muup Jy~beam^{-1}}$.}
\label{fig:lofar_widefield_zoom}
\end{figure*}

Appendix~\ref{appendix:data_calibration} details the data calibration and imaging procedures. Specifically, Appendix~\ref{sec:hba_data} describes the LOFAR-VLBI wide-field process, with the final map shown in Fig.~\ref{fig:lofar_widefield} (with a zoom on the central region of the cluster in Fig.~\ref{fig:lofar_widefield_zoom}). The procedures for the uGMRT and VLA, with the latter including both total intensity and polarization, are covered in Appendix~\ref{sec:ugmrt} and Appendix~\ref{sec:vla}, respectively, with their final maps presented in Fig.~\ref{fig:a2255_ugmrt_vlaconcat}.

\section{Methods}
\label{sec:methods}
\subsection{Spectral index maps}
\label{sec:spidx_maps}
To study the formation and characteristics of the observed filaments, we produced spectral index maps of the cluster center combining maps at 144~MHz, 1260~MHz, and 1520~MHz, having a common resolution of $1.5^{\prime\prime}$ (corresponding to a physical scale of 2.3 kpc at the cluster redshift). Highly resolved spectral index maps are important for disentangling different emitting components and substructures, especially along the Original TRG. The spectral index $\alpha$ is defined as
\begin{equation}
\label{eq:spidx}
\alpha = - \frac{\ln{(S_1/S_2)}}{\ln{(\nu_1/\nu_2)}}~,
\end{equation}
with $S_1$ and $S_2$ being the flux densities at frequencies $\nu_1$ and $\nu_2$, respectively. The errors on the spectral index values are computed using the standard error propagation formula
\begin{equation}
\label{eq:spidx_error}
\Delta \alpha = \Biggl| \frac{1}{\ln{(\frac{\nu_1}{\nu_2})}} \Biggr| \sqrt{\Bigl( \frac{\Delta S_1}{S_1} \Bigr)^2 + \Bigl( \frac{\Delta S_2}{S_2} \Bigr)^2 }~,
\end{equation}
where $\Delta S_1$ and $\Delta S_2$ are the flux density errors for $S_1$ and $S_2$, respectively, evaluated as
\begin{equation}
\Delta S = \sqrt{(f_{\rm{cal}} \cdot S)^2 + (\sqrt{N_{\rm{beams}}} \cdot \sigma_{\rm{rms}})^2}~.
\label{eq:fluxdens_err}
\end{equation}
Here, $f_{\rm{cal}}$ is the absolute flux density calibration uncertainty, $S$ the flux density, $\sigma_{\rm{rms}}$ the rms noise, and $N_{\rm{beam}}$ is the number of beams covering the source. We assumed the absolute flux density uncertainty to be $10\%$ for the LOFAR-VLBI~\citep{shimwell2022,timmerman2022}, $5\%$ for the uGMRT~\citep[as in][]{lal2020}, and $4\%$ for the VLA~\citep[as in][]{Perley13}.
\par To ensure recovery of flux on the same spatial scales for all the observed frequencies, imaging was done with an inner $uv$-cut at $420\lambda$, corresponding to the largest minimum baseline among the three interferometers (namely the uGMRT); we used a common resolution, and Briggs \texttt{robust} of $-1.5$ for the LOFAR-VLBI and $-0.5$ for the uGMRT and VLA map, to balance the short spacings coverage and recover the emission from the extended features at higher frequencies. We checked the astrometric accuracy of the uGMRT and VLA maps with respect to LOFAR-VLBI (for which we checked the astrometry of the delay calibrator), and corrected the flux density scale of the uGMRT map following the routine explained in Appendix~\ref{appendix:flux_density_scale}. This ensured the flux scale between the~\citet{scaife2012} and the~\citet{perley2017} catalogs for 3C286 to be consistent, within the chosen flux density uncertainties; for LOFAR-VLBI, the two flux scales for 3C48 were already consistent within the flux density uncertainties~\citep{perley2017}. After re-gridding to a common pixelation, we computed the spectral index for each pixel having surface brightness values above a $3\sigma_{\rm{rms}}$ threshold using \textsc{BRATS}~\citep{harwood2013,harwood2015}. This tool uses the linear least squares regression method in logarithmic space to fit the spectral index to the observed flux densities. The final spectral index map for the Original TRG is shown in Fig.~\ref{fig:spidx} (with the spectral index error map shown in Appendix~\ref{appendix:spidx_error}) and discussed in Sect.~\ref{sec:originaltrg}.

\begin{figure}[h!]
\centering
\includegraphics[width=\columnwidth]{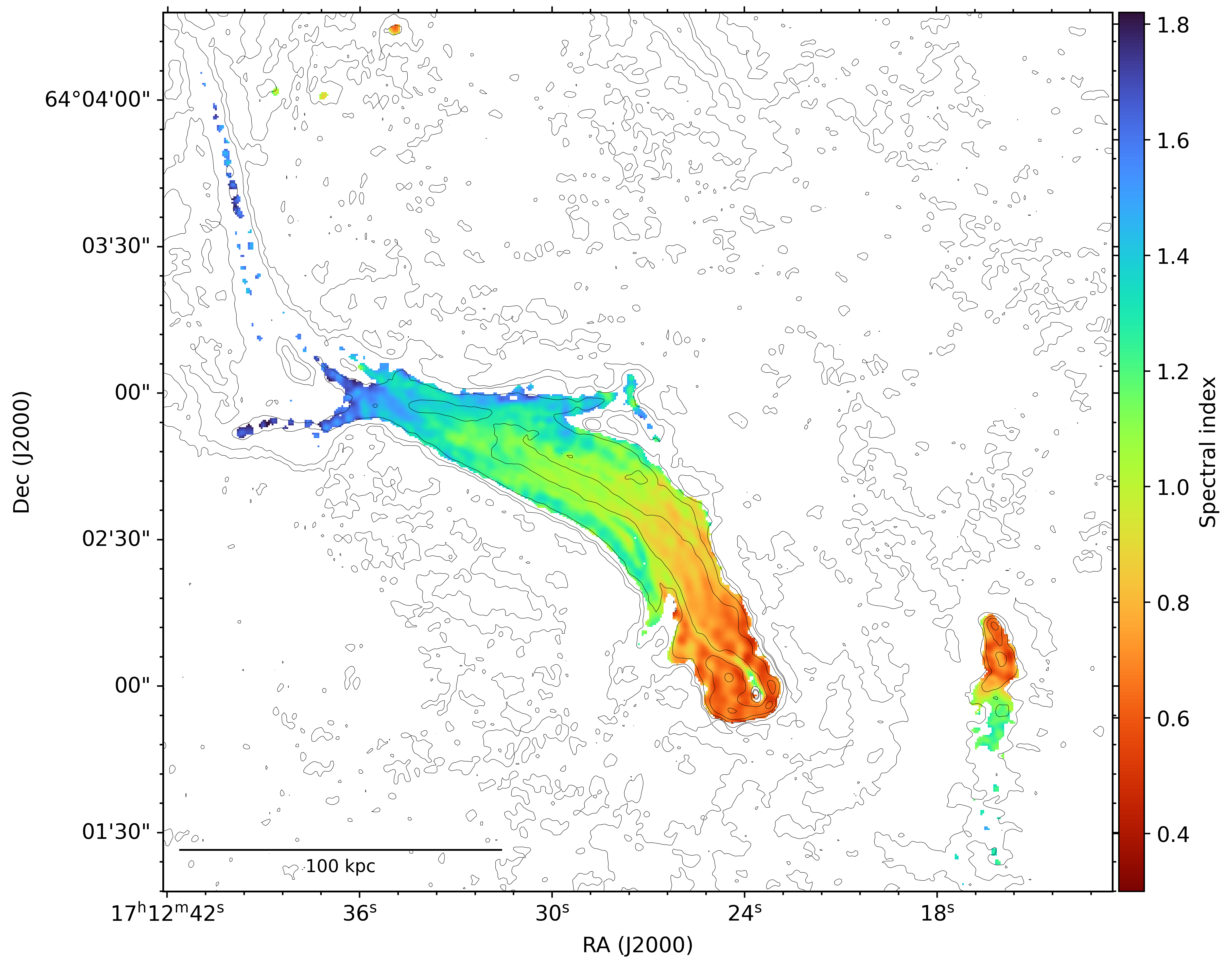}
\caption{Spectral index map using LOFAR-VLBI (144~MHz), uGMRT (1260~MHz), and VLA (1520~MHz) data for the Original TRG. Grey contours represent $3,~6,~12,~24\ldots\sigma_{\rm{rms}}$ levels, with $\sigma_{\rm{rms}} = 38~\rm{\muup Jy~beam^{-1}}$ being the noise from the LOFAR-VLBI, wide-field image at $1.5^{\prime\prime}$ resolution in Fig.~\ref{fig:lofar_widefield_zoom}.}
\label{fig:spidx}
\end{figure}

\subsection{Radiative aging}
\label{sec:radiativeaging}
Spectral index information is useful to recover the radiative age of the electrons in the tail and the filaments. Relativistic particles ejected from the core of radio galaxies are expected to progressively lose their energy mainly through synchrotron and inverse Compton radiation mechanisms. The age due to synchrotron plus inverse Compton losses is then expressed, in Myr, by the following relation~\citep{1980miley}
\begin{equation}
\label{eq:radage}
t = 1590 \cdot \frac{\sqrt{B^{\prime}_{\rm{eq}}}}{B^{\prime 2}_{\rm{eq}} + B^{2}_{\rm{CMB}}} \cdot \sqrt{\frac{1}{\nu_{\rm{b}} (1 + z)}}~,
\end{equation}
where $\nu_{\rm{b}}$ is the break frequency in GHz, $B^{\prime}_{\rm{eq}}$ is the revised equipartition magnetic field in $\rm{\muup G}$, and $B_{\rm{CMB}}$ is the equivalent magnetic field of the cosmic microwave background in $\rm{\muup G}$ as $B_{\rm{CMB}} = 3.25 (1 + z)^{2}$. Although our frequency coverage is not sufficient to determine the break frequency, we can use the formula derived in~\citep{bruno2019} to produce a radiative age map of the Original TRG using the spectral index information, the injection index, and the revised equipartition magnetic field
\begin{equation}
\label{eq:radagebruno}
t = \frac{1590}{\sqrt{(1 + z)}} \cdot \frac{\sqrt{B^{\prime}_{\rm{eq}}}}{B^{\prime 2}_{\rm{eq}} + B^{2}_{\rm{CMB}}} \cdot \sqrt{\frac{\ln{(\nu_1)} - \ln{(\nu_2)}}{\nu_1 - \nu_2} \cdot (\alpha - \alpha_{\rm{inj}})}~.
\end{equation}
To obtain the revised equipartition magnetic field, we start from calculating the \enquote{standard} equipartition magnetic field $B_{\rm{eq}}$ assuming minimum energy density in the source
\begin{equation}
\label{eq:equipartition}
B_{\rm{eq}} = \Bigl(\frac{24\pi}{7} u_{\rm{min}} \Bigr)^{1/2}~,
\end{equation}
with $u_{\rm{min}}$ being the minimum energy density in units of $\rm{erg~cm^{-3}}$ and derived as
\begin{equation}
\label{eq:umin}
u_{\rm{min}} = \xi(\alpha,\nu_{1},\nu_{2}) (1+k)^{4/7}(\nu_{0})^{4\alpha/7}(1+z)^{(12+4\alpha)/7}\times I_{0}^{4/7}d^{-4/7}.
\end{equation}
Here, $\xi(\alpha,\nu_{1},\nu_{2})$ is a constant~\citep[see Tab.~1 in][]{2004govoni}, $k$ is the proton-to-electron energy ratio, $I_0$ is the surface brightness in units of $\rm{mJy~arcsec^{-2}}$ at frequency $\nu_0$ in MHz, $z$ is the redshift of the source, and $d$ is the depth of the source in units of kpc. A filling factor of 1 was assumed. In this method, the synchrotron luminosity is calculated assuming a low and a high frequency cutoff: however, given that they depend on the energy of the emitting electrons, it is more appropriate to set a low and a high energy cutoff~\citep{brunetti1997,2005beck}. This provides the so-called \enquote{revised} equipartition magnetic field
\begin{equation}
\label{eq:revised_beq}
B^{'}_{\rm{eq}} \sim 1.1~\gamma_{\rm{min}}^{\frac{1-2\alpha}{3+\alpha}}~B_{\rm{eq}}^{\frac{7}{2(3+\alpha)}}~,
\end{equation}
where $\gamma_{\rm{min}}$ is the minimum Lorentz factor for the electrons, assuming $\gamma_{\rm{min}} \ll \gamma_{\rm{max}}$. The value of $B^{'}_{\rm{eq}}$ is based on several assumptions. $k$ depends on the mechanism that generates the emitting electrons: in the literature it is usually assumed to be $k=1$ or $k=0$~\citep{2004govoni} and this former value has been assumed for the filaments. We also assumed the filaments and the main tail to be cylinders, with depth $d$ corresponding to their projected width, and for $\alpha$ we used the integrated values corresponding to each feature combining the LOFAR and VLA maps. The lower energy cutoff $\gamma_{\rm{min}}$ is difficult to estimate because it depends on the exact shape of the radio spectrum: we assumed $\gamma_{\rm{min}} = 100$~\citep{vanweeren2009}. All these assumptions resulted in equipartition magnetic fields of $10~\rm{\muup G}$ for the main tail, $13~\rm{\muup G}$ for F1, $12~\rm{\muup G}$ for the horizontal filament and $17~\rm{\muup G}$ for the vertical one. The injection index was obtained using \textsc{BRATS}, which fits the spectral index of the target source and returns the value of $\alpha_{\rm{inj}}$ that minimizes the distribution of $\chi^2$ for the fitted model. We derived $\alpha_{\rm{inj}} = 0.55 \pm 0.01$, in agreement with the values measured close to the core of the Original TRG.
\par Using the revised equipartition magnetic field and the injection index in Eq.~\ref{eq:radagebruno}, we found radiative ages of $33-35$~Myr for F1, ranging from 37 to 44~Myr from west to east for the horizontal filament, and $\sim 36$~Myr for the vertical one. For the Trail, considering the constraints on the spectral index, the radiative age is over 33~Myr. Of course, these values are based on the value of the assumed revised equipartition magnetic field, which, as already discussed above, relies on several assumptions. These actually depend also on the full history of the fields in which electrons were found, as well as unaccounted for possible adiabatic losses. Still, these time scales are well within the typical synchrotron radiative lifetimes of electrons; moreover, compared to the dynamical lifetime of the filaments (from Paper I, $\tau_{\rm{dyn}} \sim 160-220$~Myr), we confirm that these filaments are younger than the time required to be dissipated by the turbulence.

\subsection{Color-color diagrams}
\label{sec:colorcolor}
In tailed radio galaxies, one would expect to observe a progressive decline in surface brightness and steepening of the radio spectrum moving away from the core, due to synchrotron and inverse Compton losses of the relativistic particles~\citep[e.g.,][]{feretti1998}. However, thanks to the high resolution, we observed multiple narrow components that bind to the tail of the Original TRG showing enhanced surface brightness and local flattening of the spectral index (Fig.~\ref{fig:spidx}). Moreover, the different filaments traced in the vicinity of the Original TRG show different spatial trends of spectral index, possibly not compatible with pure aging scenario (Sect.~\ref{sec:originaltrg}). This motivates a better investigation of different aging models, to constrain the nature and possibly the formation of these features. Radio color-color diagrams, first introduced by~\citet{katzstone1993}, represent a useful tool to trace the spectral shape of different regions of a source. These are obtained by comparison of local spectral indices computed at two pairs of frequencies (specifically, the x-axis and y-axis report the low-frequency and high-frequency spectral index, respectively). The points located on the bisector correspond to regions of plasma with a power-law spectrum having injection spectral index $\alpha_{\rm{inj}}$; the ones located below correspond, instead, to regions showing particles aging ($\alpha > \alpha_{\rm{inj}}$). Such diagrams are generally used to discriminate between theoretical synchrotron aging spectral models. In fact, the position of the points depends on the aging model: the most common are the Jaffe-Perola~\citep[JP,][]{jaffe1973}, the Kardashev-Pacholczyk~\citep[KP,][]{kardashev1962,pacholczyk1970}, and the Tribble-Jaffe-Perola~\citep[TJP,][]{tribble1993}. They are all characterized by a single injection event that produces a power-law distribution of relativistic electrons. Both JP and KP models assume a uniform magnetic field. For the JP model, the pitch angles of the synchrotron-emitting electrons are continuously isotropized on timescales shorter than the radiative timescale. For the KP, instead, the pitch angle remains in its original orientation with respect to the magnetic field direction. The TJP model is built on the JP one, introducing Gaussian spatial fluctuations of the magnetic field around a central value ($B_0$). For our analysis, we used a low-frequency spectral index between LOFAR-VLBI at 144~MHz and uGMRT at 1260~MHz ($\alpha_{144}^{1260}$) and a high-frequency spectral index between uGMRT at 1260~MHz and VLA at 1520~MHz ($\alpha_{1260}^{1520}$). In this case, the spectral index and the associated error were calculated using Eq.~\ref{eq:spidx} and~\ref{eq:spidx_error}. We used \textsc{BRATS} to calculate the synchrotron spectrum produced by the different aging models, using the revised equipartition magnetic field and the injection index from Sect.~\ref{sec:radiativeaging}. The resulting color-color diagrams for the main tail and the filaments are shown in Fig.~\ref{fig:colorcolor}, and are discussed in Sect.~\ref{sec:originaltrg}.

\subsection{Rotation measure synthesis}
\label{sec:rmsynth}
VLA data in A configuration were calibrated for polarization (Sect.~\ref{sec:vla}) and used for the analysis of the Original TRG and its filaments. Given their faintness, and the disadvantaged position of the radio galaxy (within the depolarizing ICM, close to the cluster center), we tried to detect polarized emission using the rotation measure (RM) synthesis technique~\citep{2005brentjens}. This technique allows combining the polarized signal of each channel that constitutes the full band ($1008-2032$~MHz in this case) to avoid bandwidth depolarization, which is an instrumental effect that occurs when a significant rotation of the polarization angle of the radiation (so-called Faraday rotation effect) is produced across the observing bandwidth and the polarization fraction is computed averaging over the band~\citep[][]{1998sokoloff}. We refer to~\citet{2005brentjens} for a detailed description of this procedure. In brief, it uses an observing bandwidth split up into many individual narrow-frequency channels. Adding up the individual channels may cause bandwidth depolarization; however, using the value of Faraday depth that maximizes the signal resulting from the co-addition of the polarized flux from all channels, it is possible to recover the polarized flux.
\par
\begin{figure*}[h!]
\centering
\includegraphics[width=\textwidth]{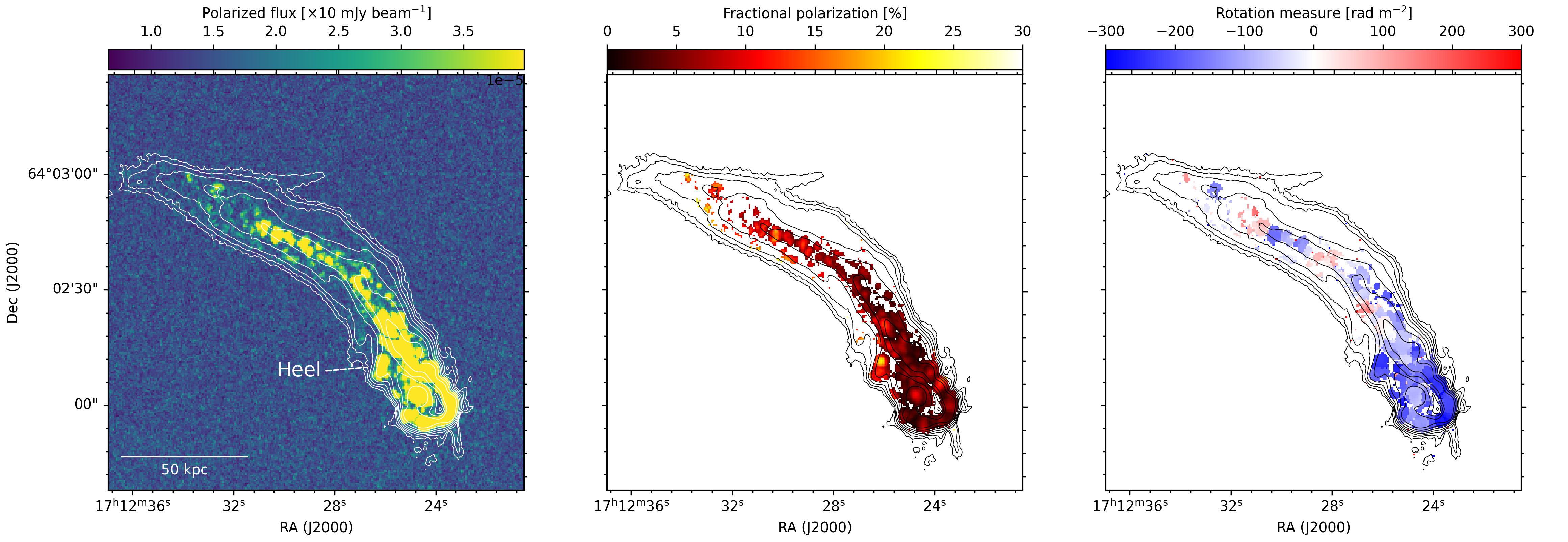}
\caption{Polarization results for the Original TRG at $2^{\prime\prime}$ resolution using the VLA. From left to right: polarized flux, fractional polarization, and rotation measure map. Contours in black are from the VLA map in total intensity used for the polarization analysis at $3, 6, 12, 24, 48, 96, 192~\sigma$ noise level, with $\sigma = 10~\rm{\muup Jy~beam^{-1}}$.}
\label{fig:polarization}
\end{figure*}

 To apply this technique, we first produced multiple deconvolved images in Stokes $Q$ and $U$ at different frequencies with \textsc{WSClean}, having a common resolution of $2^{\prime\prime}$ (corresponding to the resolution of the lower frequency channel): these cubes are required by the \textsc{RM-Tools~v1.4.8}~\citep{2020rmtools}, which is the software used for the RM synthesis. To have enough signal per channel, required to detect also the more extended emission in tails and filaments, we produced 48 images across the entire band, with bandwidth of 16~MHz each. This also ensures that, within each channel, we are sensitive to a maximum observable Faraday depth of $~655~\rm{rad~m^{-2}}$, and so to avoid significant bandwidth depolarization within each channel, since the Faraday depth in non-cool-core clusters is generally below $500~\rm{rad~m^{-2}}$~\citep{2016bohringer}. We used the \textsc{RMsynth3D} tool of the \textsc{RM-Tools} software to process the Stokes $Q$ and $U$ cubes and deconcolved the Faraday spectrum with \textsc{RMclean3D}: the final map, which contains the maximum polarized intensity in each pixel, was then used for polarization analysis. We corrected for the Ricean bias, following~\citet{2012george}, evaluating the polarization noise at the outer boundaries (in the ranges [$\mp 1000$,$\mp600$]~$\rm{rad~m^{-2}}$) of the Faraday spectrum for each pixel. We corrected the RM for the Galactic foreground using a value of $37.8~\rm{rad~m^{-2}}$~\citep{hutschenreuter2022} and assuming it to be constant over the radio galaxy. In Fig.~\ref{fig:polarization} the polarized intensity, the polarization fraction, and the RM map are shown.

\section{Results}
\label{sec:results}
\begin{figure}[ht!]
\centering
\includegraphics[width=0.8\columnwidth]{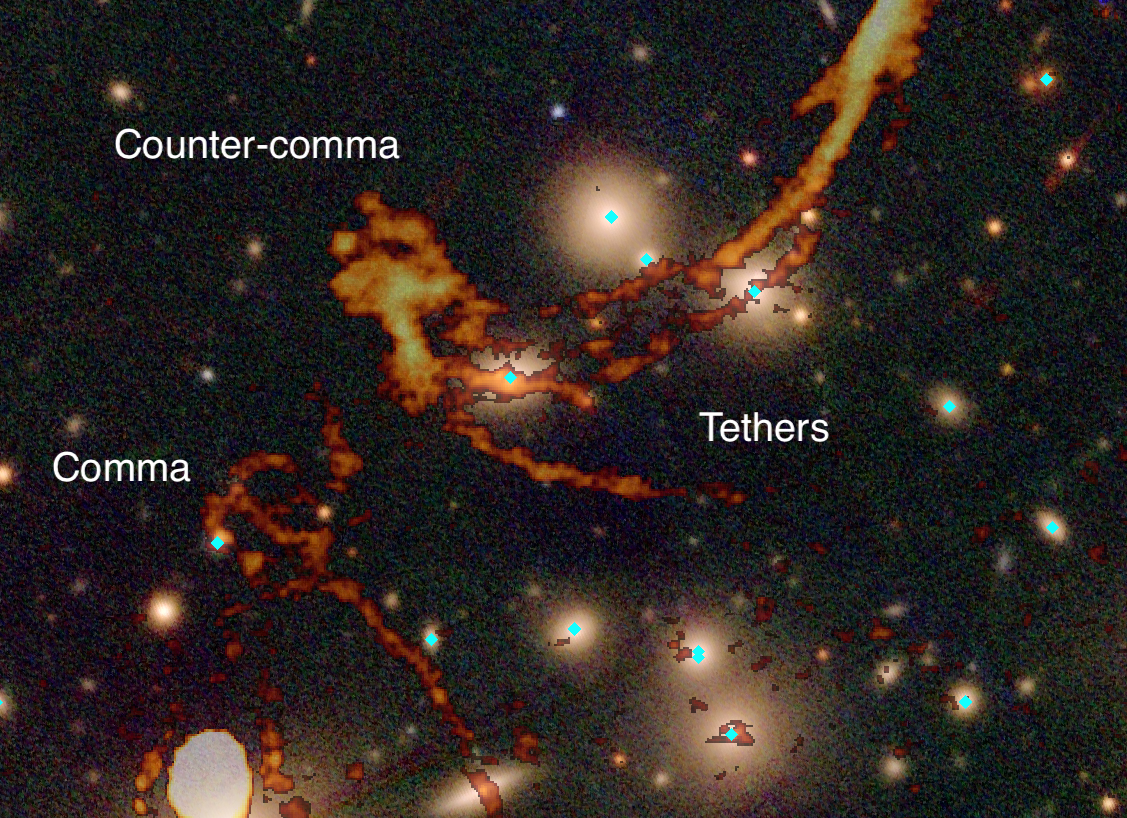}\\
\includegraphics[width=0.8\columnwidth]{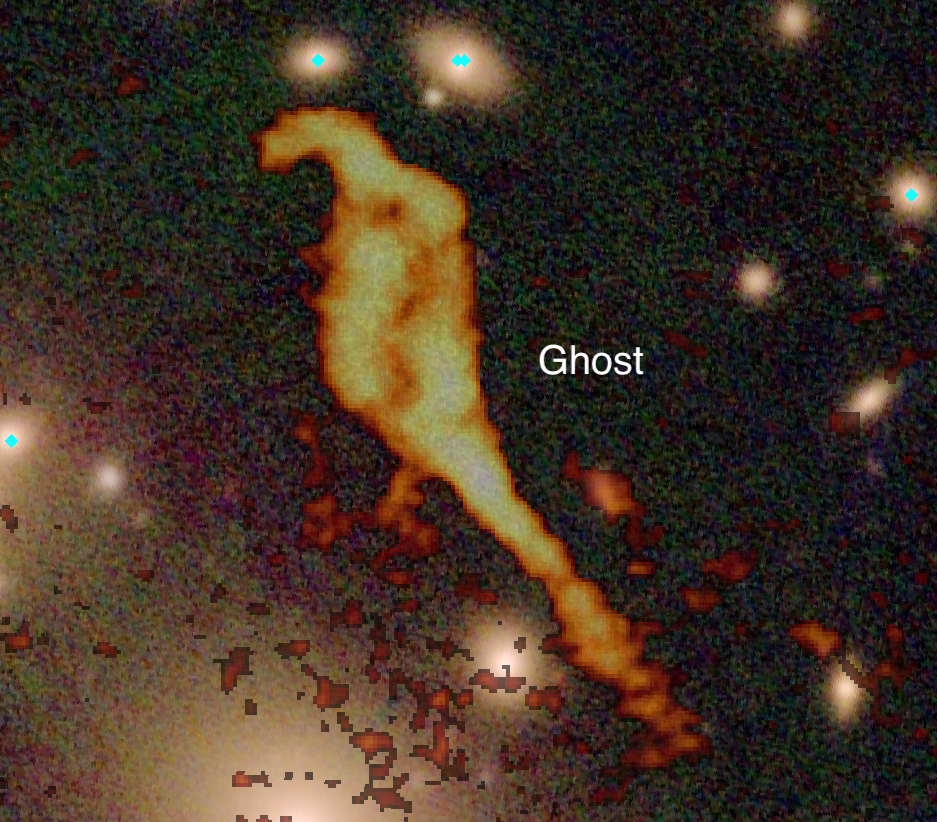}\\
\includegraphics[width=0.8\columnwidth]{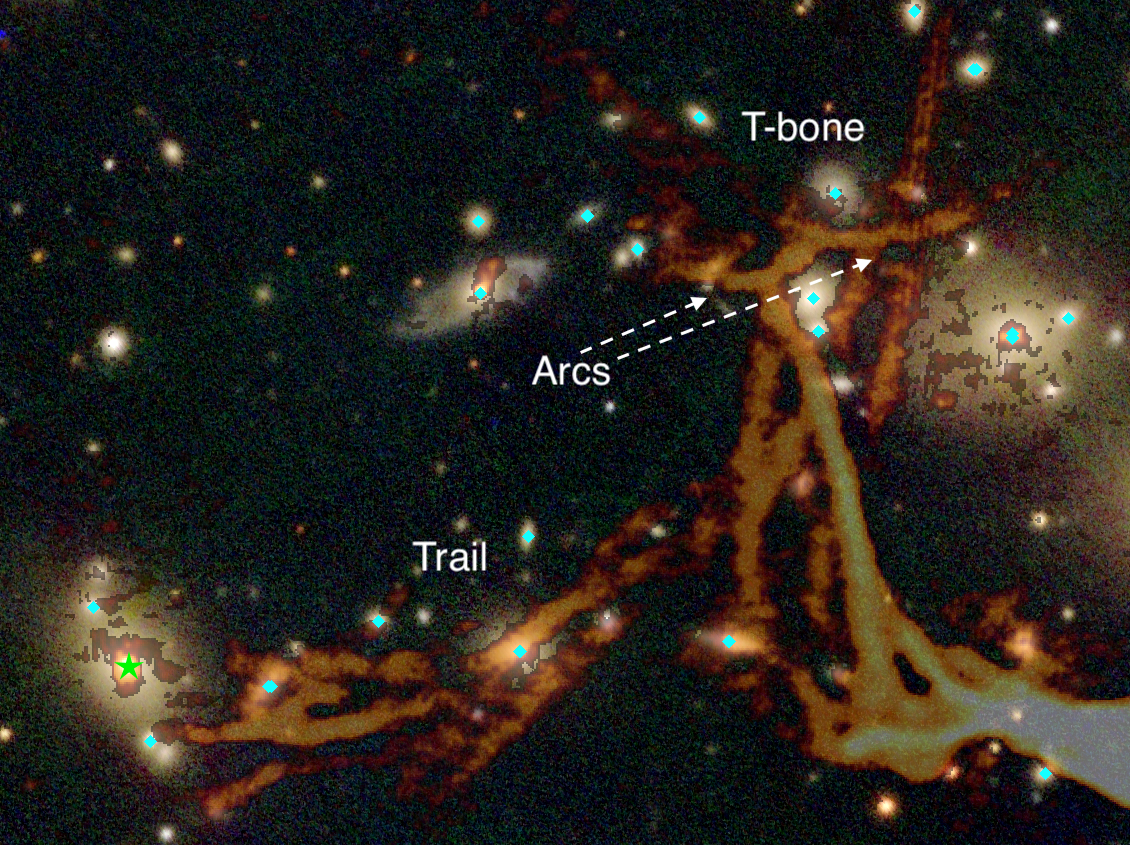}
\caption{Radio features observed from the LOFAR-VLBI image in Fig.~\ref{fig:lofar_widefield}. Radio emission is shown in orange/yellow, superimposed on an optical image from the Panoramic Survey Telescope \& Rapid Response System~\citep[PanSTARRS,][]{panstarrs2016}. Cyan rhombi represent cluster member galaxies from~\citet{yuan2003},~\citet{shim2011}, and~\citet{tyler2014}, which may have provided electrons to generate these features. Top: Comma, Counter-comma, and tethers. Center: the Ghost. Bottom: the Trail and T-bone.}
\label{fig:optical}
\end{figure}
\begin{table}
\centering
\caption{Main properties of the filaments and diffuse patches of radio emission discussed in this paper.}
\label{tab:filaments}
\begin{tabular}{ccccc}
\hline\hline
Name & Length & $\langle\rm{Width}\rangle$ & $S_{\rm{144~MHz}}$ & $\alpha$\\
 & [kpc] & [kpc] & [mJy] & \\
\hline\hline
Comma & 154 & 7 & $45 \pm 5$ & $> 2.3$\\
C-comma & 147 & 7 & $110 \pm 10$ & $> 2.3$\\
Ghost & 107 & $9.5$ & $100 \pm 10$ & $>2.5$\\
Trail & 250 & $7.5$ & $404 \pm 40$ & $>2.0$\\
T-bone & 85 & 7.5 & $78 \pm 8$ & $> 2.0$\\  
Horizontal & 110 & 9 & $319 \pm 32$ & $1.41 \pm 0.05$\\
Vertical & 105 & 7 & $210 \pm 21$ & $1.74 \pm 0.05$\\
F1 & 83 & 5 & $168 \pm 17$ & $1.13 \pm 0.05$\\
F2 & 60 & 6 & $154 \pm 15$ & $1.12 \pm 0.05$\\
\hline\hline
\end{tabular}
\tablefoot{Column 1: name of the feature from Paper I and from~\citet{botteon2020}, reported in Fig.~\ref{fig:lofar_widefield_zoom} and Fig.~\ref{fig:lofar_widefield}. Column 2: projected length in kpc from Fig.~\ref{fig:lofar_widefield}. Column 3: average width in kpc from Fig.~\ref{fig:lofar_widefield}, evaluated as in Paper I. Column 4: flux density in mJy at 144~MHz from Fig.~\ref{fig:lofar_widefield} with flux density error as in Eq.~\ref{eq:fluxdens_err}. Column 5: integrated spectral index between 144~MHz and 1520~MHz.}
\end{table}
In this section, we detail the characteristics and results obtained for each of the main features of the cluster. First, we briefly describe the results for features that were only detected in the LOFAR-VLBI image, and then we focus on the main target of this investigation, the Original TRG. The lengths reported are projected, while the widths are obtained following the procedure already described in Paper I. The results for the main features are listed in Tab.~\ref{tab:filaments}. In Fig.~\ref{fig:trends} we also report the surface brightness (left panel) and spectral index (right panel) trends along some of the main features of the Original TRG and the filaments (regions in Fig.~\ref{fig:trends_regions}).

\subsection{Comma, Counter-comma, and tethers}
\label{sec:commas}
The Comma and Counter-comma (C-comma in Tab.~\ref{tab:filaments}) are two quasi-specular patches located between the end of the Goldfish's tail and the Double, discovered by~\citet{botteon2020} and labeled in Fig.~\ref{fig:lofar_widefield}. The former is closer to the Double and has a flux density at 144~MHz, measured within $3\sigma_{\rm{rms}}$ level, of $45 \pm 5$~mJy and a length of $\sim 154$~kpc; the latter, instead, has a flux density of $110 \pm 10$~mJy and a length of $\sim 147$~kpc. Both features are undetected at higher frequencies with the current sensitivity of our uGMRT and VLA observations, which constrain their spectrum to be steeper than $\alpha \sim 2.3$. They also have a similar morphology, presenting a \enquote{head} in the upper part ($\sim 35$~kpc wide), followed by a thin curved component which extends below it (length of $120$~kpc for the Comma, and$\sim 95$ kpc for the Counter-comma). The Counter-comma is attached directly to the bottom end of the Goldfish's tail (Fig.~\ref{fig:optical}) still at high-resolution, suggesting a possible connection between the two; this, however, can also be due to projection effects. There are multiple thin components connecting the Counter-comma and the Goldfish resembling the tethers observed in the Mystail in Abell 3266~\citep{rudnick2021} and in the radio galaxy IC711 in Abell 1314~\citep{vanweeren2021}. The A2255 tethers are made of three components, with widths ranging between $4-8$~kpc. Their origin is still unclear: there are three cluster member galaxies observed in the optical~\citep{yuan2003, tyler2014} corresponding to the position of the tethers (two superimposed, one just above, highlighted in cyan in Fig.~\ref{fig:optical}) which could have provided electrons to the system. The Comma upper end is spatially coherent with two potential host galaxies, one of these identified as a cluster member~\citep[$z=0.0809$,][]{yuan2003}, that could have provided electrons to supply the observed emission.

\subsection{Ghost}
\label{sec:ghost}
The Ghost is another radio patch discovered by~\citet{botteon2020}. It is well-detected at 144~MHz ($\sim 0.10$~Jy, $5\sigma$ level, flux density) and has a length of 107 kpc and a width up to 34 kpc. At high-resolution (with LOFAR-VLBI) it appears to consist of two distinct components of $9-10$~kpc each. At the Ghost position there is no optical counterpart (middle panel of Fig.~\ref{fig:optical}), which makes any interpretation about the origin of this patch of emission non-trivial. One possibility is that it represents residual radio emission from past activity of a galaxy that moved away from its original position; it can also be emitting plasma which moved away from its host galaxy because of the ongoing cluster merger. As already seen for the Comma and Counter-comma, the non-detection of the Ghost at higher frequencies allowed us to constrain its spectrum to be $\alpha > 2.5$~\citep[in agreement with][]{botteon2020}.

\subsection{Trail and T-bone}
\label{sec:trail}
Together with the filaments directly attached to the Original TRG, discovered in Paper I, there are other bundles of filaments located, in projection, beyond 200 kpc from the radio galaxy core. These have been labeled the Trail and were discovered by~\citet{botteon2020}. They extend for about 250 kpc in projection, with a radius of curvature of 180 kpc. These filaments are not observed at higher frequencies by the uGMRT and the VLA, resulting in a very steep spectrum~\citep[$\alpha = 2-2.5$,][]{botteon2020}. Thanks to LOFAR-VLBI wide-field calibration and imaging, we resolved many narrow extended components in the Trail at $1.5^{\prime\prime}$. These filaments are less bright than the others in the Original TRG: the Trail at 144~MHz is about 15-25 times less bright than the main tail, 6-15 times than the horizontal filament and 2-3 times than the vertical filament. The Trail may represent the remnant of past AGN radio activity from a nearby host-galaxy, as suggested also by the presence of a cluster member galaxy at its bottom-left end (RA: $\rm{17^h 13^m 02^s.04}$, Dec: $64^{\circ} 03' 08^{\prime\prime}.3$, highlighted with a green star in the bottom panel of Fig.~\ref{fig:optical}). At high-resolution, we observe the Trail to be detached from this candidate host galaxy. Considering also the lack of surface brightness trends along this feature~\citep[as shown by the red line in Fig.~\ref{fig:trends} and claimed also in][]{botteon2020}, usually observed along aging tails, it is plausible that the plasma in the Trail was injected by a previous activity of the host candidate and then went to some re-energization process, possibly related to the turbulent cluster center. High-resolution spectral index studies would be required to confirm this scenario, but with current facilities there is no way to trace such features at higher frequencies with matching resolution given their steep spectral index.
\par The T-bone is a quasi-horizontal narrow component of around 85 kpc length and 7 kpc width which \enquote{terminates} both the Trail and the vertical filament. As already observed in other cases, going to high-resolution we resolved for the first time two arc-like structures (Fig.~\ref{fig:optical}, bottom panel), with a radius of curvature of 20 and 55 kpc.

\subsection{Original TRG}
\label{sec:originaltrg}
\begin{figure*}[h!]
\centering
\includegraphics[width=0.4\hsize]{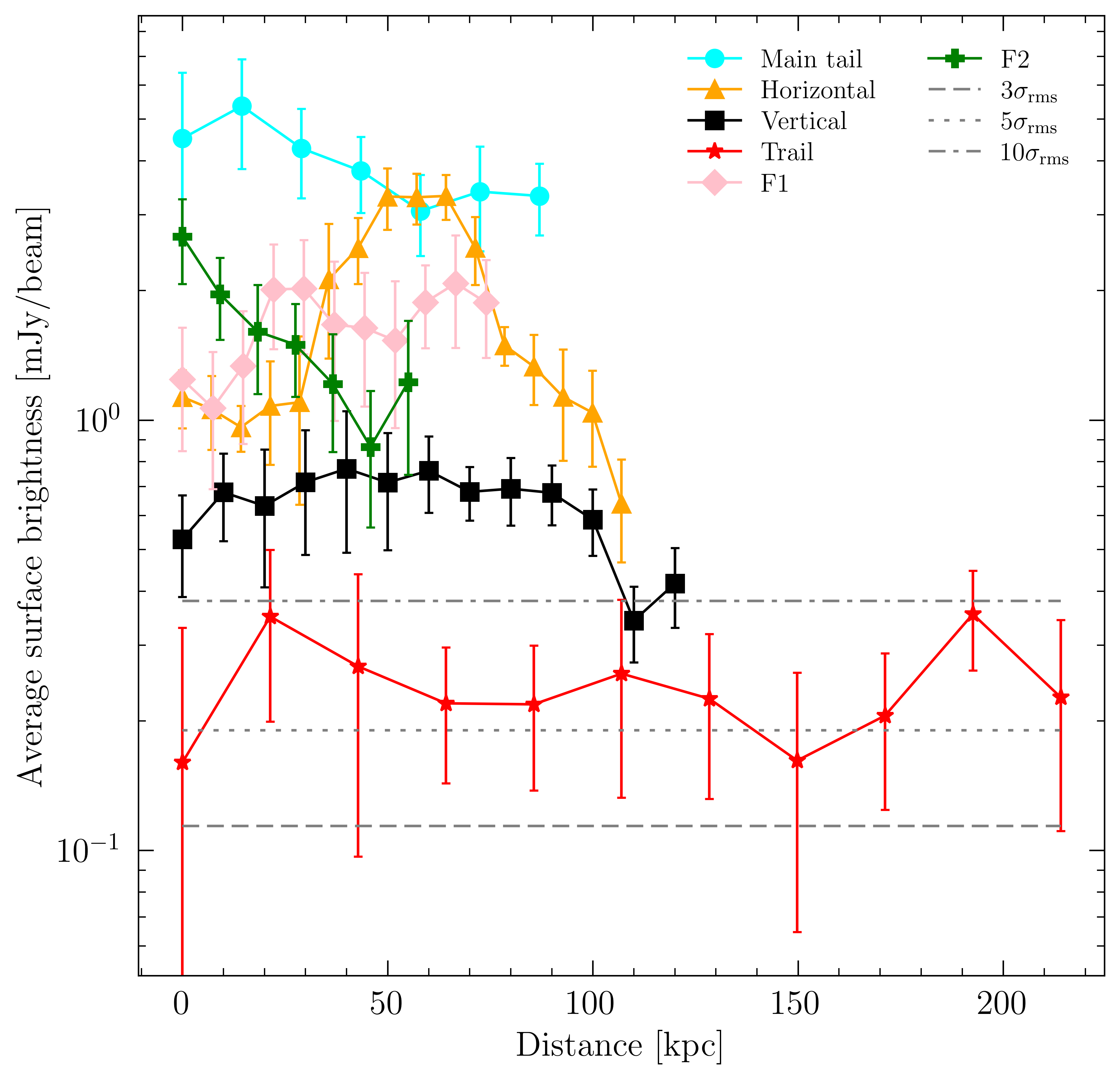}\quad
\includegraphics[width=0.4\hsize]{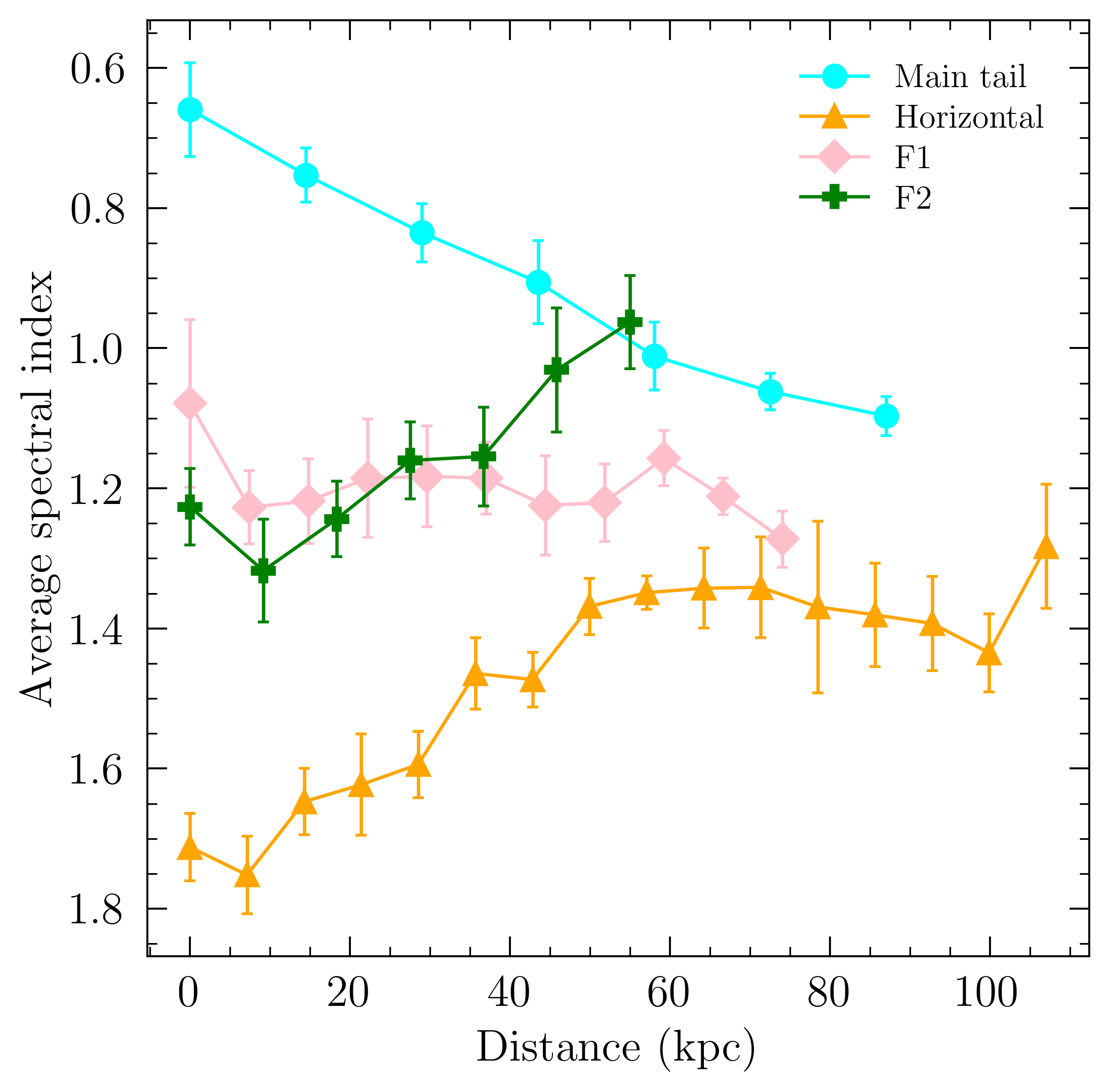}
\caption{Surface brightness and spectral index trends for the Original TRG and the related filaments. The regions used to evaluate the average surface brightness and spectral index, with different color corresponding to different features, are shown in Fig.~\ref{fig:trends_regions}, as well as the arrows that indicate the direction of the trends. These are the main tail (cyan), the horizontal (orange) and vertical (black) filaments, F1 (pink), F2 (green), and the Trail (red). Left: Average surface brightness trends along the features' extension at 144~MHz. Right: Average spectral index trends along the features' extension, considering only pixels above a $3\sigma_{\rm{rms}}$ threshold. The spectral index is not shown for the vertical filament and for the Trail given the lack of enough pixels above the chosen threshold.}
\label{fig:trends}
\end{figure*}
The Original TRG is the brightest tailed radio galaxy in A2255, well detected also at higher frequencies and showing multiple filamentary substructures. For this reason, it was designated as the main focus of our investigation. It is a NAT extending for $\sim 270$ kpc overall, considering all its components. In Paper I, we discovered and detailed many filamentary features observed at sub-arcsecond resolution along its tail: in particular, we discovered the presence of two filamentary extensions alongside the main tail, namely F1 and F2, a horizontal filament at the upper end of the tail, and a vertical filament extending upward (Fig.~\ref{fig:lofar_widefield_zoom}). These same features are also observed at higher frequencies with the uGMRT and VLA (Fig.~\ref{fig:a2255_ugmrt_vlaconcat}). Being at lower resolution (with respect to Paper I), we detected the filaments along their entire extent, as well as other features which were barely detected at high-resolution, such as the fork at the western end of the horizontal filament or the gap observed between the southern jet and F1 (Fig.~\ref{fig:lofar_widefield_zoom}). The morphological properties of the filaments in the Original TRG, listed in Paper I, are also reported in Tab.~\ref{tab:filaments}.
\begin{comment}
\begin{figure}[h!]
\centering
\includegraphics[width=\columnwidth]{integrated_spidx.png}
\caption{Integrated spectrum of the main features of the Original TRG between 144~MHz and 1520~MHz. The dashed lines show the fitted power law.}
\label{fig:integratedspectrum}
\end{figure}
\end{comment}
\par Spectral index values from 144~MHz to 1520~MHz range from 0.5, close to the host galaxy, to $\sim1.7-1.8$ on the vertical filament and the left side of the horizontal one (Fig.~\ref{fig:spidx}). The upper jet of the TRG, which then develops into the main tail, shows spectral steepening along its extension, in line with plasma aging under the assumption of pure radiative losses due to synchrotron and inverse Compton emission after relativistic particles are ejected from the core of the radio galaxy~\citep[e.g.,][]{cuciti2018,bruno2024,koribalski2024}. This approximately continuous steepening along the main tail (Fig.~\ref{fig:trends}, right panel, in cyan) proceeds for about 142 kpc, before it links with the horizontal filament. There is local flattening of the spectrum corresponding to the eddies that develop from the northern jet, possibly due to the instabilities driven in the jet by the interaction with the ambient medium~\citep[e.g.,][]{odea1987,sakelliou1999}. F1 and F2 show a substantially different spectral behavior along their extension (Fig.~\ref{fig:trends}, right panel, F1 in pink and F2 in green). The spectral index remains almost constant for F1, meaning that the relativistic plasma that constitutes these features has a different energetic history with respect to the one in the main tail. Moreover, spectral index values in F1 suggest that, as already proposed in Paper I, it is not associated with the \enquote{missing} southern jet: after the gap, there is, in fact, a spectral index discontinuity between what comes from the southern jet ($\alpha = 0.8$) and values at the bottom of F1 ($\alpha = 1.3$), which does not follow the same aging spatial trend shown by the main tail. Moreover, spectral index values along the F1 do not correspond to those of the main tail at the corresponding distance from the core. There is also a component (right above the main tail, highlighted in Fig.~\ref{fig:spidx}) which shows spectral index values compatible with the main tail but morphologically not coincident with it. It may either reflect the large spectral complexity which characterizes the tail, or represent some component which is hidden behind the main tail such as a filament or, potentially, the missing tail from the southern jet, which may have been bent behind the main tail in projection. The horizontal filament shows spectral flattening moving from east (almost 1.8) to west (around 1.4). The spectral index difference between the two sides, together with the enhanced brightness corresponding to the central part of the filament (already highlighted in Paper I), suggests the superimposition of multiple components, one corresponding to the steeper left side and one coming from the upper side, connected to the vertical filament, with a flatter spectral index; otherwise, the flatter part of the filament could come from the relativistic plasma deposited by the radio galaxy that underwent localized re-acceleration. From the spectral index map, only a few pixels have a detection above the $3\sigma_{\rm{rms}}$ for the vertical filament. We also produced integrated spectra for the main features of the Original TRG, combining the flux density measurements from 144~MHz to 1.52~GHz. The resulting spectra are reported in Tab.~\ref{tab:filaments}. The entire tail, comprehending also the southern jet and termination, has a spectral slope of $0.69 \pm 0.05$. The filaments are steeper, with F1 of $1.13 \pm 0.05$, F2 of $1.12 \pm 0.05$, the horizontal of $1.41 \pm 0.05$, and vertical of $1.74 \pm 0.05$.
\par
\begin{figure}[h!]
\centering
\includegraphics[width=\columnwidth]{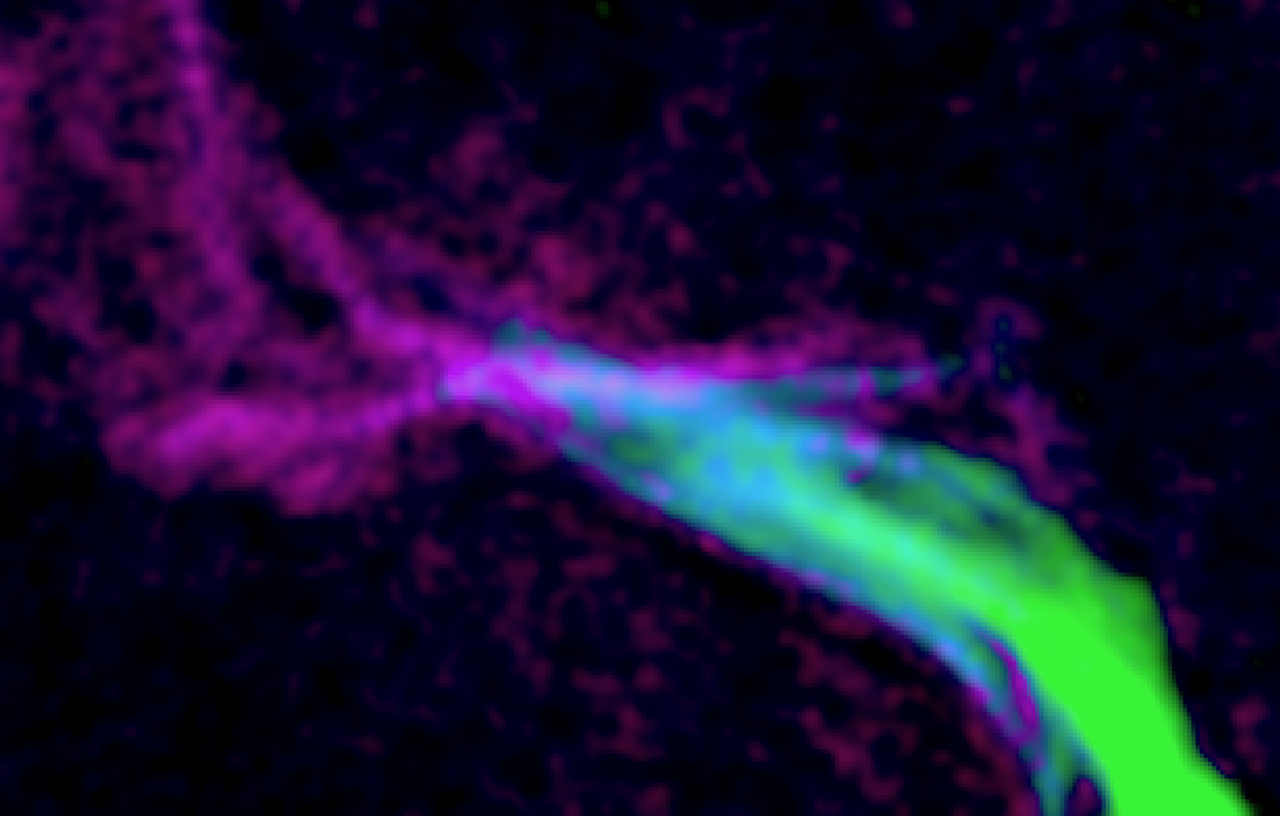}
\caption{Spectral index tomography map between 144~MHz and 1520~MHz, where different colors highlight different spectral ranges. $\alpha < 1.2$ in green, $1.2<\alpha<1.6$ in cyan, $\alpha > 1.6$ in magenta.}
\label{fig:sptomomaps}
\end{figure}

 The different spectral components in the Original TRG are better highlighted in Fig.~\ref{fig:sptomomaps}, where different colors identify different spectral index ranges. We notice how both F2 and the horizontal filament are rather formed by multiple, overlapping components. The bottom side of F2 seems rather connected to the tail electron component, supporting the fact that it is probably part of the southern bent jet emerging behind the main tail. The horizontal filament is also formed by at least two components, with $\alpha = 1.6$ being a good discriminant between the two. The steeper component (Fig.~\ref{fig:sptomomaps}, in magenta) occupies mainly the eastern part of the filament, with a thin extension towards the west just above the main body of the filament. The eastern and brightest part of the horizontal filament (Fig.~\ref{fig:sptomomaps}, in cyan) is instead flatter. Part of it seems to move upwards, towards the vertical filament, and also towards F2: disentangling all these components is however non-trivial, even at this resolution. These two spectral components overlap around the point where the central part of the horizontal filament connects with the main tail. This spectral index bimodality for the horizontal filament confirms what is already observed from the trends in Fig.~\ref{fig:trends} (right panel, in orange).

\begin{figure*}[ht!]
\centering
\subfloat{\includegraphics[width=0.36\textwidth]{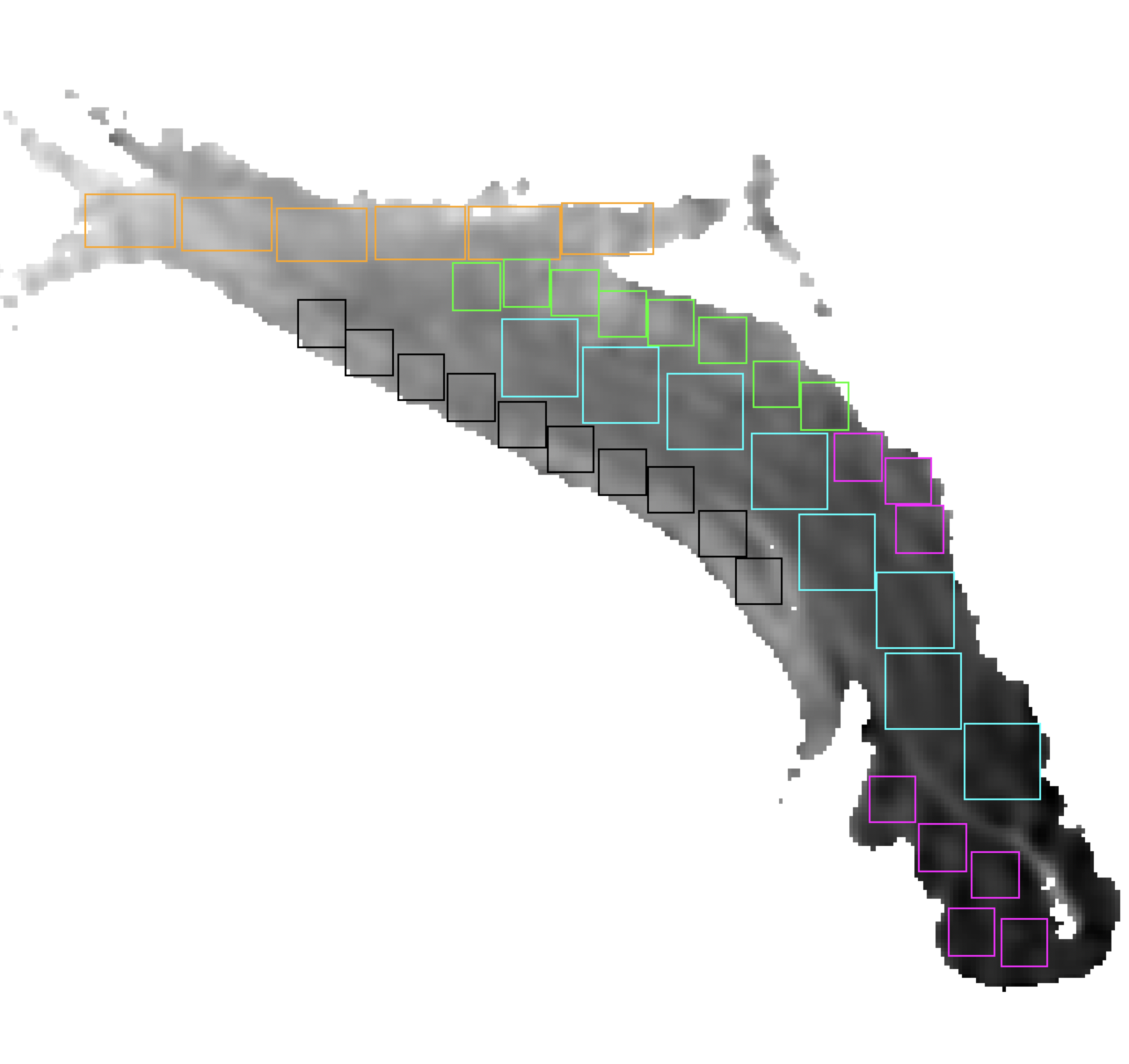}}
\subfloat{\includegraphics[width=0.36\textwidth]{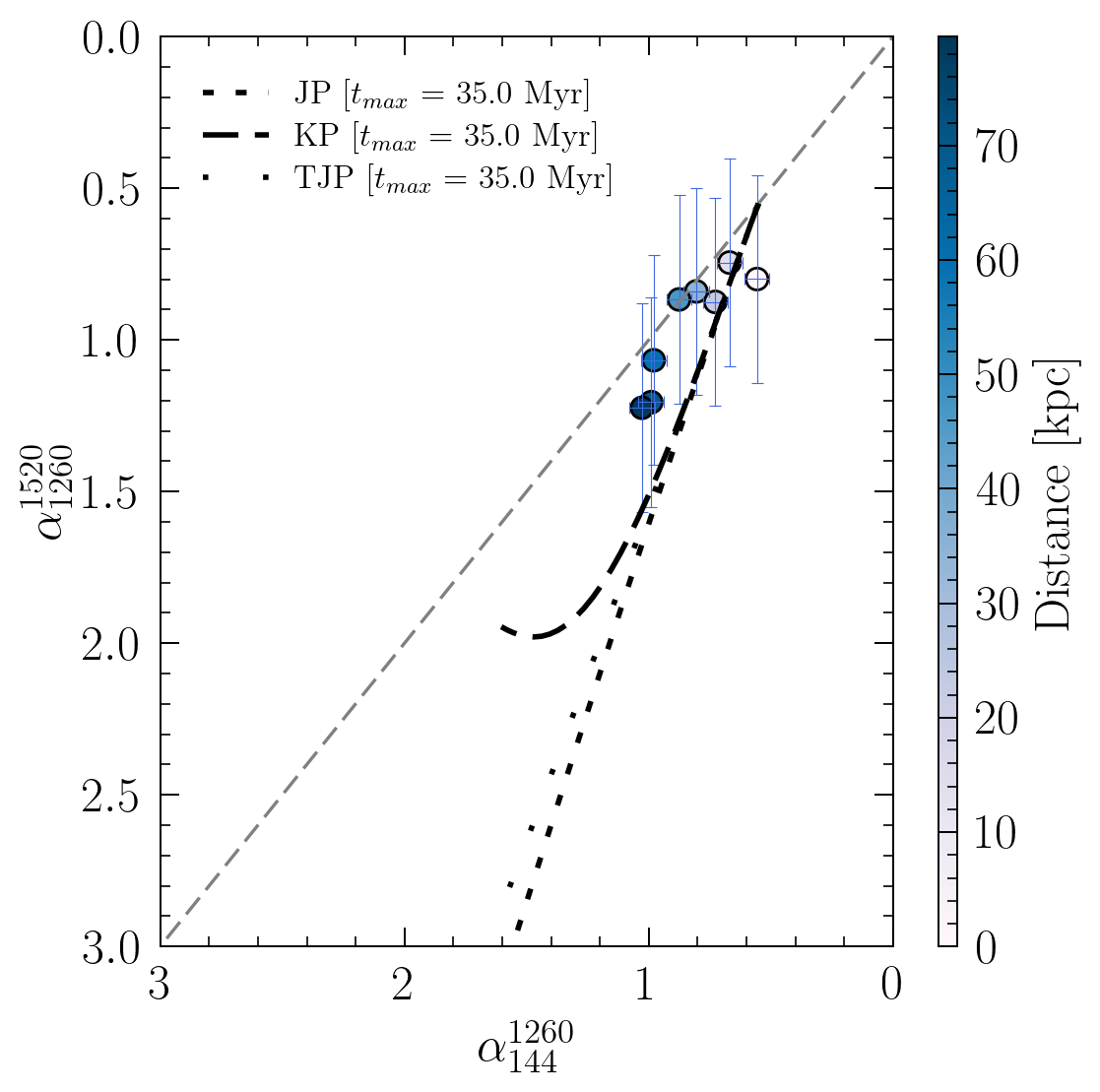}}\\
\subfloat{\includegraphics[width=0.36\textwidth]{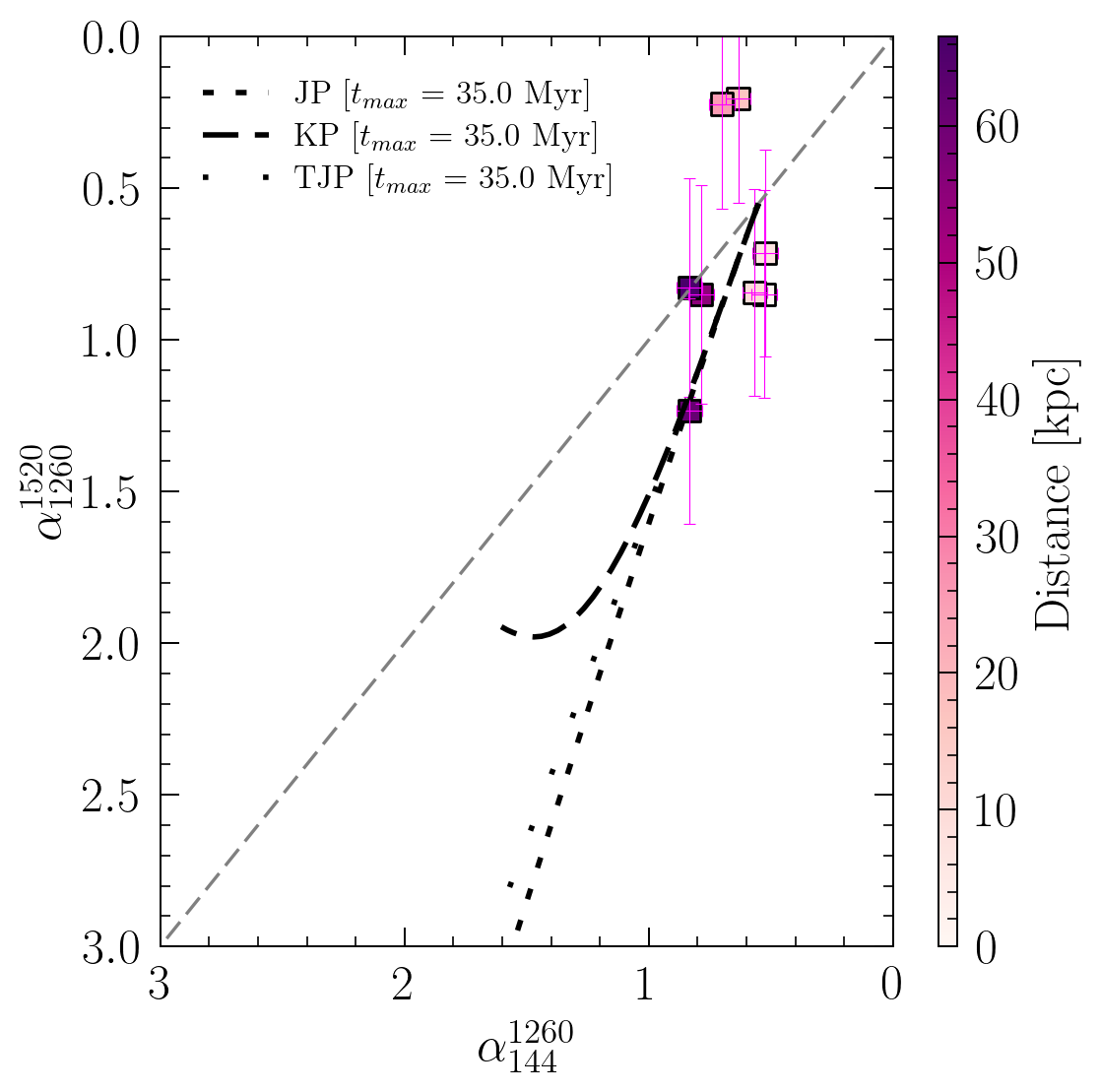}}
\subfloat{\includegraphics[width=0.36\textwidth]{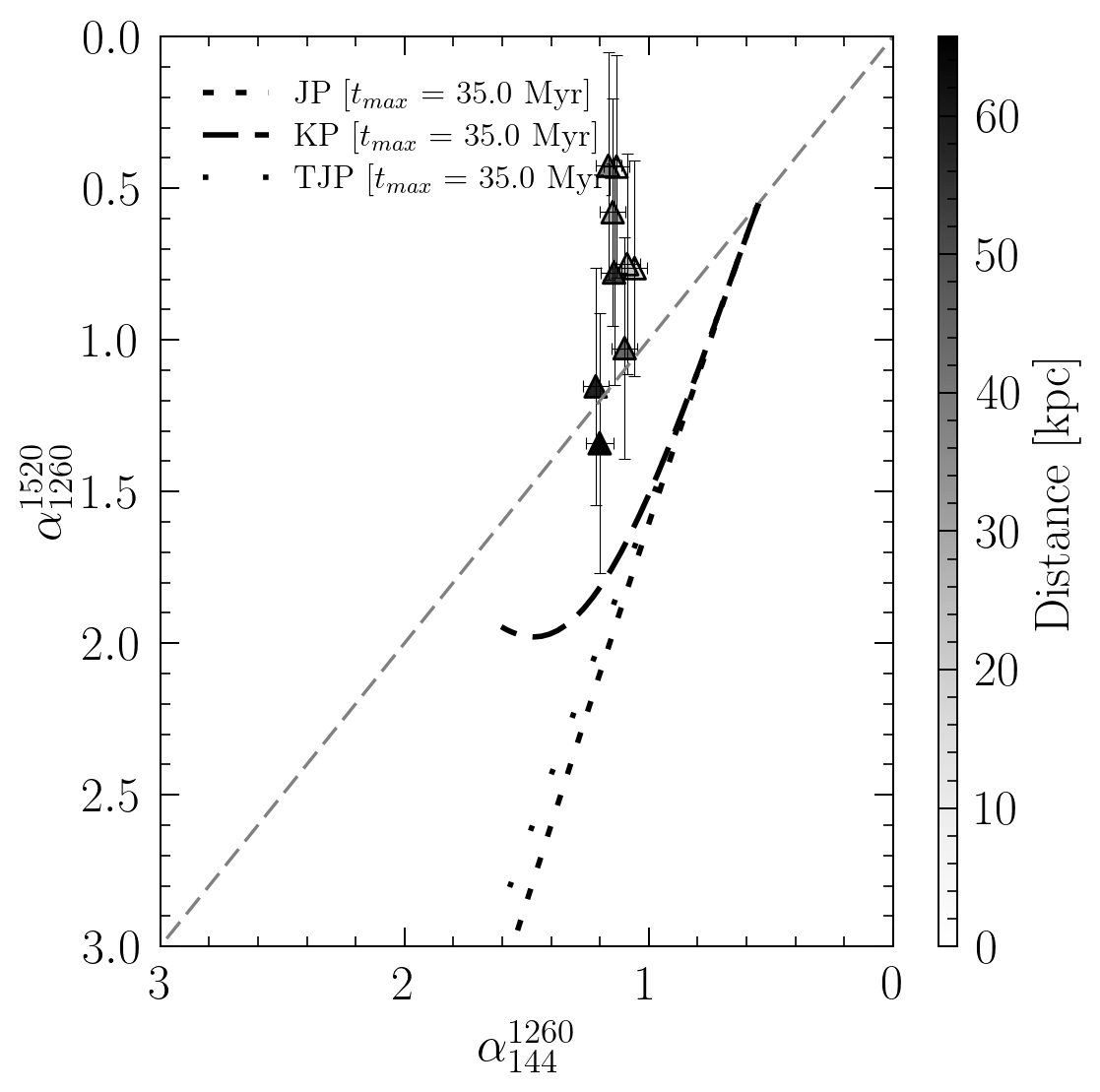}}\\
\subfloat{\includegraphics[width=0.36\textwidth]{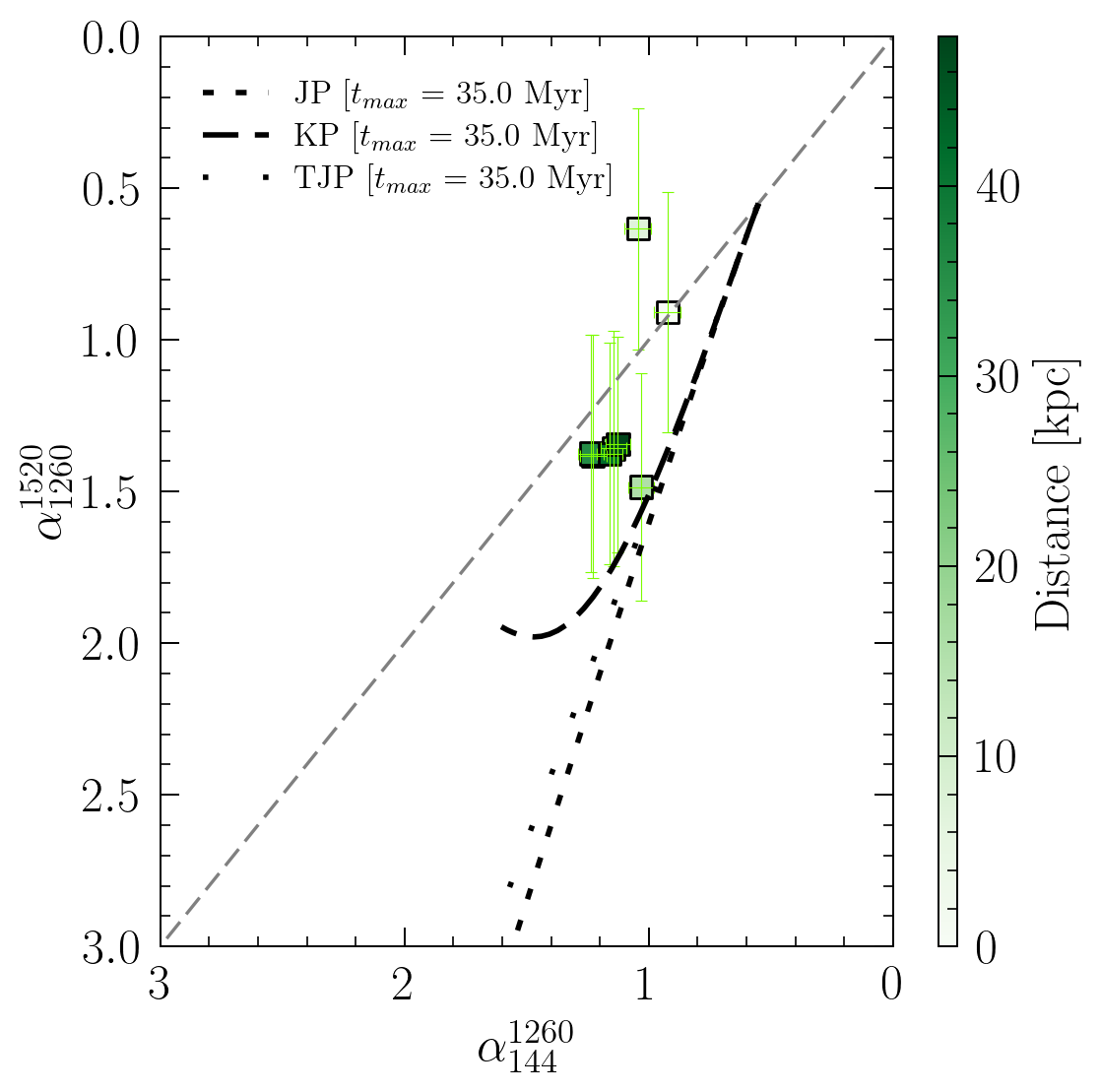}}
\subfloat{\includegraphics[width=0.36\textwidth]{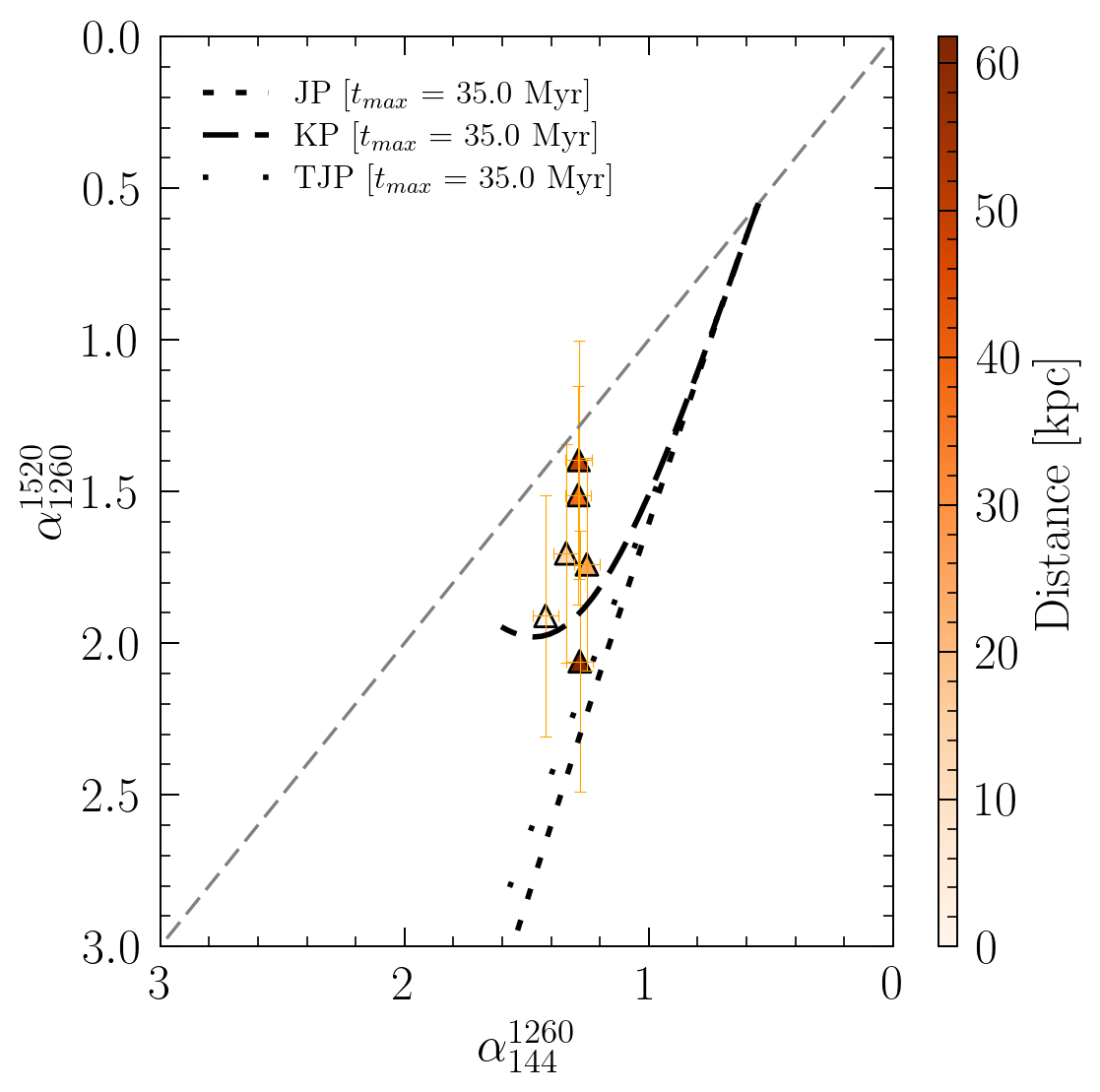}}
\caption{Color-color diagrams for different regions in the Original TRG. From top to bottom, from left to right: main tail (blue), southern tail (magenta), F1 (black), F2 (green), horizontal filament (orange). The distance is calculated between the centers of consecutive regions, starting from a reference region (Sect.~\ref{sec:originaltrg}), in the left-right direction for the horizontal filament, and in the bottom-top direction for all the others.}
\label{fig:colorcolor}
\end{figure*}
\par Color-color diagrams for the main features of the Original TRG are shown in Fig.~\ref{fig:colorcolor}. The large complexity of the structure and the number of features of interest required us to isolate these on the color-color diagrams to better constrain their spectral properties. Each data point corresponds to the value of the spectral index within a region, with the color scale changing depending on the distance of each region from a \enquote{reference}: this is assumed to be the region with lower declination, apart from the horizontal filament for which it is the one on the eastern end of the feature. Lines with different styles trace the ideal spectral index trend determined by different aging models (Sect.~\ref{sec:colorcolor}). The large number of details revealed at high-resolution makes the interpretation of these plots non-trivial. 
Following the downstream deployment of the northern jet into the main tail (blue data points), the spectral index follows the aging trend predicted by all the tested models. However, the point are rather close to the power-law line, suggesting that although the aging appears monotonic, it is not due to a single electron population, but some mixture. Following the implications from the spectral tomography (Fig.~\ref{fig:sptomomaps}), we divided the analysis of the filament originally labeled as F2 into two components. The magenta data points represent the bottom part, which, as stated before, can possibly be related to an additional component obscured by the main tail. Connecting it with the emission coming from the southern jet, we observe how the trend is somewhat similar to the one found for the main tail, apart from the very first data points closer to the core. However, the possible co-existence of different electron components within the same defined areas prevents a clear conclusion from being drawn. Green data points are instead from the top part of F2, and show a similar behavior apart from the top-eastern part of F2 which presents spectral index values very close to each other. The straighter filaments, namely F1 (in black) and the horizontal one (in orange), show instead a different and peculiar trend. Both show a small spread at low-frequency ($1.0-1.3$ for F1, $1.2-1.5$ for the horizontal, within the errors), varying then significantly at high-frequency ($0.0-2.3$ for F1, $1.0-3.0$ for the horizontal). 
\par The interpretation of these values for the filaments is not obvious. They are generally not consistent with a single electron population, but rather with a great deal of inhomogeneity within the boxes in terms of magnetic fields, or electron populations, or both. The electron populations could differ based on different acceleration sites, or losses. Despite the fact that filaments appear as coherent distinct structures, each with their own spectral trends, the color-color diagrams show that they are each complex mixtures, possibly due to unresolved substructures such as bundles of smaller filaments. From the available observations, it is not possible to unambiguously distinguish all the mixed emitting components.

\par Polarization results are shown in Fig.~\ref{fig:polarization}. This source was already observed to be polarized at $5~\rm{GHz}$ and $8~\rm{GHz}$, having a mean fractional polarization between $12.5-14\%$ with fractional polarization increasing from the core ($2-4\%$) along the tail~\citep{govoni2006}. At lower frequencies, the source was reported to have a fractional polarization of $2.1\%$, $2.0\%$, and $1.4\%$ at 18, 21, and 25 cm, respectively~\citep{pizzo2011}. Thanks to the combination of high sensitivity and resolution, we recovered polarized emission from the head of the Original TRG, as well as along the brightest part of the tail up to the data point at which it connects to the horizontal filament. We report fractional polarization values at $1.5$~GHz of $\sim 4-5\%$ at the core, which increase to $6-7.5\%$ at the location of the northern jets. The inner regions of the radio source are expected to be more depolarized being more deeply embedded in the host galaxy~\citep{osullivan2015}. The southern jet, instead, shows even lower values ($2-3\%$): this difference is also found in the RM map, with the northern jet displaying an average RM of $-214~\rm{rad~m^{-2}}$ and the southern of $-268~\rm{rad~m^{-2}}$. The polarization fraction along the tail is not uniform but rather patchy, with local enhancements corresponding to the turbulent eddies (up to $13\%$) and close to the connection with the horizontal filament (up to $18-22\%$). On average, moving upward along the tail, values increase from $4\%$ to $9\%$. Together with some pixels for the horizontal filament, we found polarized emission for F1 up to $22\%$ in small patches in the middle and upper part of the filament. Interestingly, there is enhanced fractional polarization corresponding to the round southern jet termination ($\sim 10\%$), as well as to the narrow component emerging from the main tail labeled the Heel in Fig.~\ref{fig:polarization} ($10-22\%$). The termination was already observed to be connected to the southern jet through thin features similar to tethers in Paper I. The Heel is then connected to the termination through a fainter component, observable both at low and high frequencies. Both components show compatible spectral properties, as well as polarization enhancement similar to that of the turbulent eddies: this may raise the possibility that they represent two components of the southern jet, which gets bent due to the instabilities that start developing in the tail, before it deploys behind the main tail. In Fig.~\ref{fig:rmdistribution} we show the distribution of RM values for the Original TRG. Large absolute values and variations in RM are expected for sources deeply embedded in the ICM, and close to the cluster center, due to the Faraday rotation that it generates on their polarized emission~\citep{feretti1998, govoni2001, 2010bonafede}. Along the tail RM values are predominantly negative, meaning that the magnetic field component along the line of sight, $B_{\rm{\parallel}}$, is mainly directed away from the observer.

\section{Discussion}
\label{sec:discussion}
In this section, we discuss the main properties reported previously in this paper for the Original TRG and its associated non-thermal filaments, addressing their morphological and spectral characteristics and their formation.
\subsection{Structure of the Original TRG and its filaments}
\label{sec:structure}
Non-thermal radio filaments have been frequently observed in recent years in different objects in the cluster environment, whether they are connected to the lobes of a radio galaxy~\citep[e.g.,][]{perley1984, maccagni2020, ramatsoku2020, condon2021,cotton2025}, involve complex interactions with jets~\citep[e.g.,][]{rudnick2022}, or are in buoyant AGN bubbles~\citep[e.g.,][]{brienza2021,giacintucci2025}. Similar structures have also been discovered within diffuse emission, such as in radio halos and mini halos~\citep[e.g., Perseus, Ophiucus, or the NE and S spurs in A2255 itself,][]{vanweeren2024,botteon2025,botteon2020} and in radio relics~\citep[e.g.,][]{owen2014, digennaro2018, rajpurohit2020, 2022rajpurohit, 2022degasperin}.
\par Thanks to the unprecedented high resolution and sensitivity at 144~MHz, we traced the morphological features of the steep and extended filaments beyond the tail, which are only observable at such low frequencies. Through our spectral studies, we have disentangled multiple spectral components along the Original TRG. The structure of the tail is, in fact, not homogeneous, with several narrow components emerging both from the total intensity and spectral index maps: this is possibly common to these sources when observed at high resolution, due to the strong instabilities developed downstream of the radio galaxy~\citep{lal2004, 2017gendron}. As already highlighted in Sect.~\ref{sec:originaltrg}, spectral and polarization information was also important to highlight the presence of a component, related to additional filaments or the missing southern tail. From the total intensity map in Fig.~\ref{fig:lofar_widefield_zoom} (as well as at sub-arcsecond resolution in Paper I), the southern tail seems to terminate abruptly in a round feature, connected through narrow tethers. There is the possibility that it may be reappearing from behind the main tail, after being obscured by this; the bending point may then correspond to the round termination. It may also be a filament (F1-like) obscured by the main tail. Being these structures thoroughly embedded into the downstream structure, is not possible to clearly separate them from the flows from the original jets. For the Heel, instead, the large polarization fraction (up to $22\%$) and RM values suggest that it may not be related to the tail, but rather some additional filamentary component arising from behind the main tail. Polarization is a useful tool also to constrain the direction of the two jets. In fact, there is a polarization fraction and RM difference between the northern and southern jet, as stated in Sect.~\ref{sec:originaltrg}. Assuming a comparable thermal electron density in both jet locations, the southern jet's larger absolute rotation could be attributed to its orientation, receding from the observer, whereas the northern jet points more towards us. This may support the evidence discussed above (and already presented in Paper I) regarding the orientation between the two jets and the consequent configuration of the tails, with the northern tail settling above the southern one. Projection effects are a primary source of ambiguity as they hinder the disentanglement of numerous overlapping components: RM analysis is a powerful tool for addressing this, despite the inherent assumptions it involves. For the brightest part of the main tail, the progressively lower absolute value of RM moving towards the end suggests that the core region is located deeper in the ICM, with the tail deploying towards the outskirts. However, there are also many spots that show higher absolute RM values or have a positive RM sign (Fig.~\ref{fig:polarization}), complicating the interpretation. Unfortunately, we miss similar information for the more outlying and faint filaments, where it would be even more important to decipher their 3D configuration to understand their relationship with the radio galaxy.
\par The configuration of the radio galaxy and filaments observed in A2255 resembles the one in Abell 3627~\citep{koribalski2024}, with a NAT radio galaxy moving within the ICM and a system of steep, curved spectra filaments ($\alpha \sim 1.9-2.5$) extending perpendicularly at the end of the tail. In A2255, we can distinguish two sets of filaments following their location, morphology, and spectral properties. F1, F2, horizontal, and vertical are all directly related to the Original TRG: they were also observed to be interconnected at higher resolution in Paper I, and show overall spectral indices between $\alpha = 0.9-1.8$ and structural linearity. The horizontal filament resembles the one observed for Source B in Abell 2443~\citep{cohen2011}. It was explained to be pre-existing filament which was re-energized by the passing of source B. The Trail and possibly the T-bone instead have steeper spectra ($\alpha > 2$) and are much fainter than the first group. Part of the Trail overlaps with the end of the vertical filament, probably due to projection effects, and also the T-bone seems to terminate the vertical filament in its upper side. No hint of curvature has been observed for the filaments with the available frequency coverage, given also the very narrow difference between the uGMRT and VLA frequencies. The spectral index map suggests how structures that seem to be uniform from the brightness maps are instead made of multiple, overlapping, electron components. This was observed through tomography in Fig.~\ref{fig:sptomomaps}, which highlights the difference between the eastern and western sides of the horizontal filament, as well as for F2 which is made by two distinct underlying components. For the filaments inside the tail of the Original TRG we did not observe any significant spectral index trend: this is particularly true for F1 and for the western side of the horizontal filament (Fig.~\ref{fig:trends}). This can be explained by the large Alfv\'en speed due to the magnetic field lines stretching (and consequent magnetic field enhancement). An example of this is the filament discovered in the radio galaxy J1333–3141 in Abell 3562, where the magnetic field is thought to be stretched behind a sloshing cold front that aligns with the radio filament~\citep{giacintucci2022}. Electrons would have to diffuse fast enough along the magnetic field lines to cover the entire filaments' extension within their radiative ages~\citep{churazov2025}. Electrons diffusing along field lines in a time $t$ can cover a distance of $d \sim 2\sqrt{D_{\parallel}t}$, where $D_{\parallel}$ is the diffusion coefficient along the field lines. Considering the radiative ages (reported in Sect.~\ref{sec:radiativeaging}) and lengths (reported in Tab.~\ref{tab:filaments}), we found values for $D_{\parallel} \sim 1.5 \times 10^{31}~\rm{cm^2~s^{-1}}$. This value is extremely large~\citep{2014brunetti}, implying super-Alfv\'enic streaming along the filaments with a scattering timescales of the electrons $\tau_{\rm{s}} \sim 1600~\rm{yr}$. Large Alfv\'enic speeds may reflect either enhanced magnetic field or ICM depletion in the filaments, which would mean that the plasma beta in these structures, as already proposed in~\citet{rudnick2022}, approaches unity. In the case of Abell 194, a spatial correlation between the radio filaments and a dip in the thermal X-ray emitting plasma was observed. The underlying idea is that the enhanced magnetic pressure may reach a significant fraction of the ambient thermal pressure, and together with accompanying cosmic rays can expel the local thermal plasma. For the Original TRG, we used the Chandra ACIS-I data used by~\citet{botteon2020} to investigate any possible correlation between the filaments and the thermal emission at high-resolution. However, the filaments are mostly located at the intersections of the detector's chips, making any interpretation about the thermal component non-trivial. Given also that deeper Chandra observations of the cluster are scheduled (460 ks, P.I. Rajpurohit), we defer this analysis for the future.

\subsection{Formation scenarios for the non-thermal filaments}
\label{sec:formation}
To explain the nature and formation scenarios of the observed filaments, we propose and discuss several scenarios.
\subsubsection{Instabilities in the downstream region}
One possibility to explain the formation of the filaments is that they arise from the instabilities generated downstream of the Original TRG, with the radio galaxy being the main provider of electrons to the filaments. The radio galaxy is crossing the denser region of the ICM, which is already turbulent due to the cluster merging state, generating additional turbulence downstream. In Paper I, we observed jet bending due to ram pressure effects close to the core. Moving upward, we resolved several regions, labeled eddies, which represent the development of dynamical instabilities that arise due to the bending and complex motions in the turbulent wake of the radio galaxy. Observation of the tail structure at sub-arcsecond resolution in Paper I confirmed the large complexity of such structures already resulting from higher frequency observations~\citep[e.g.,][]{lane2002, owen2014, 2020gendron} and simulations~\citep{oneill2019a, ohmura2023}. These phenomena are possibly common to any tailed radio galaxy when observed at sufficiently high resolution. The radio galaxy can contribute in two different ways to the formation of the filaments. Crossing the ICM, it deposits fresh electrons in its wake that may be subject to both the turbulence generated by the merger and the crossing. Turbulence generates magnetic field lines stretching/bending, enhancing the magnetic field and so the corresponding synchrotron emission for the electrons, illuminating the filaments.
\par Another possibility is that the radio galaxy \enquote{injects} the filaments in its wake due to its interplay with the ICM. In Paper I, we resolved filaments in the envelope of the tail, such as F1 and, partially, F2. The dynamical instabilities at the boundary of the tail can strip these filaments and mix them in the ICM motions behind. In this scenario, we would observe different stages of this mechanism at different resolutions. At sub-arcsecond resolution, we observed the raising of the instabilities at the beginning of the tail in the forms of eddies and bending of the tail. These create the filaments observed around the main tail, which are at the first stage of their evolution. Then, at $1.5^{\prime\prime}$ (Fig.~\ref{fig:lofar_widefield_zoom} and~\ref{fig:a2255_ugmrt_vlaconcat}), we can see distant filaments that were once part of the radio galaxy and now exist independently. These are the horizontal and vertical filaments (closer, flatter and brighter), and the Trail (further, steeper, and fainter). At the very end of this process, the filaments will feed the ICM with plasma that is unable to emit significant radio emission but as soon as it is efficiently mixed and re-energized by other mechanisms it can provide seed electrons to the diffuse radio structures observed in this cluster at lower resolution~\citep[such as the giant radio halo or radio relics,][]{botteon2020, botteon2022}.
\par The lack of spectral index trends in the filaments inside the radio galaxy, as highlighted in Fig.~\ref{fig:trends} and Sect.~\ref{sec:structure}, further supports the main role of the radio galaxy. The filaments part of the Trail, left behind the radio galaxy, are instead expected to show spectral steepening because the fresh plasma extracted from the Original TRG is mixed in a turbulent medium (turbulent ICM plus turbulence generated by the radio galaxy crossing), being subject to expansion and/or compression and resulting in steepening of the spectrum. Indeed, we observe different spectral values for the filaments in the Trail, which are steeper than $\alpha = 2$.

\section{Conclusions}
\label{sec:conclusions}
Non-thermal filaments represent new intriguing phenomena for studying the complex magnetic and particle processes that coexist in the ICM. In this work, following the detections already presented in Paper I, we analyzed the main spectral and polarization properties of the filaments in the central region of the galaxy cluster A2255. We obtained a deep, wide-field ($1.5^{\circ} \times 1.5^{\circ}$) map of the cluster at $1.5^{\prime\prime}$ using LOFAR-VLBI which, combined with new and archival uGMRT and VLA data at higher frequencies, allowed for a high resolution spectral index analysis of the filaments, crucial to disentangle the multitude of sub-structures and try to constrain their formation scenario. VLA data were also calibrated for polarization, expected with large fractional polarization in presence of ordered magnetic fields. We summarize hereafter the main findings of this paper.
\begin{enumerate}
\item In LOFAR-VLBI map, we detected a bundle of extended ($\sim250~\rm{kpc}$) and steep spectrum ($\alpha > 2$) filaments, known as the Trail and the T-bone, placed (in projection) at the end of the Original TRG. The high resolution allowed us to disentangle multiple components in these structures, with widths ranging between $4-11$~kpc. 
\item We found integrated spectral index values for the filaments of 0.69 (main tail), 1.13 (F1), 1.12 (F2), 1.41 (horizontal), and 1.74 (vertical).
\item From VLA data, we recovered fractional polarization up to $22\%$ in a few patches of F1, increasing up to $18-22 \%$ moving upward along the main tail.
\item Spectral index analysis revealed that F2 is rather formed by two overlapping components: one going downward, and the other coming up from the horizontal filament. Also, the horizontal filament resulted to be not homogeneous, with the flatter western side which appears to be connected to a component coming down from the vertical filament.
\item Overall, there is signature of the presence of an additional component obscured by the main tail from the spectral index and polarization information, which emerges from the main tail. We speculate that it may be the southern tail which re-emerges after being bent behind what we see as the main tail. However, the large complexity of the structures, which deeply overlap, hinders a clear detection of this potential tail.
\end{enumerate}
From their spectral and morphological properties, we believe that the filaments directly related to the Original TRG are low-$\beta$ plasma structures, illuminated by enhanced magnetic fields due to line stretching. For the Trail, we miss clear spectral information, but the very steep spectral index $\alpha > 2$ suggests that its plasma originates from the Original TRG. We propose that the radio galaxy has a crucial role in the formation of the filaments. Such structures can arise from the plasma deposited by the jets' activity which is then subject to turbulence injected by the motion of the host galaxy added to the one related to the cluster merger. Another possibility is that the filaments are stripped away directly from the radio galaxies because of instabilities in the downstream region. We resolved the starting point of such instabilities in Paper I, the eddies: these then turn into filaments, and the filaments are successively dissipated by the turbulence, providing seed electrons to the ICM which can then be re-energized to generate large-scale diffuse radio emission~\citep{botteon2020, botteon2022}.
\par Deep, upcoming X-ray observations are critical to confirm the low-$\beta$ plasma scenario, with potential X-ray dips serving as key indicators at the filament locations. Additionally, we have around 300 hours of overall cluster data available for calibration and imaging. Processing these data can benefit from recently-developed strategies, like sidereal visibility averaging~\citep[][]{dejong2025sidereal}, which efficiently reduce the computational expense of imaging.

\begin{acknowledgements}
We thank the anonymous referee for the helpful and constructive comments on this manuscript. We thank Annalisa Bonafede and Marco Balboni for the valuable discussions and suggestions that helped improving the contents of the paper. EDR and CG are supported by the Fondazione ICSC, Spoke 3 Astrophysics and Cosmos Observations. National Recovery and Resilience Plan (Piano Nazionale di Ripresa e Resilienza, PNRR) Project ID CN\_00000013 \enquote{Italian Research Center for High-Performance Computing, Big Data and Quantum Computing} funded by MUR Missione 4 Componente 2 Investimento 1.4: Potenziamento strutture di ricerca e creazione di \enquote{campioni nazionali di R\&S (M4C2-19)} - Next Generation EU (NGEU). MB acknowledges support from INAF under the following funding schemes: Large Grant 2022 (project \enquote{MeerKAT and LOFAR Team up: a Unique Radio Window on Galaxy/AGN co-Evolution}) and Large GO 2024 (project \enquote{MeerKAT and Euclid Team up: Exploring the galaxy-halo connection at cosmic noon}). JMGHJdJ acknowledges support from project CORTEX (NWA.1160.18.316) of research programme NWA-ORC, which is (partly) financed by the Dutch Research Council (NWO), and support from the OSCARS project, which has received funding from the European Commission’s Horizon Europe Research and Innovation programme under grant agreement No. 101129751. ELE is grateful for support from the Medical Research Council [MR/T042842/1]. LKM is grateful for support from a UKRI FLF [MR/Y020405/1] and LOFAR-UK through STFC [ST/V002406/1]. This manuscript is based on data obtained with the International LOFAR Telescope (ILT). LOFAR~\citep{vanhaarlem2013} is the Low Frequency Array designed and constructed by ASTRON. It has observing, data processing, and data storage facilities in several countries, which are owned by various parties (each with their own funding sources), and which are collectively operated by the ILT foundation under a joint scientific policy. The ILT resources have benefited from the following recent major funding sources: CNRS-INSU, Observatoire de Paris and Université d’Orléans, France; BMBF, MIWF-NRW, MPG, Germany; Science Foundation Ireland (SFI), Department of Business, Enterprise and Innovation (DBEI), Ireland; NWO, The Netherlands; The Science and Technology Facilities Council, UK; Ministry of Science and Higher Education, Poland; The Istituto Nazionale di Astrofisica (INAF), Italy. This research made use of the LOFAR-IT computing infrastructure supported and operated by INAF, including the resources within the PLEIADI special \enquote{LOFAR} project by USC-C of INAF, and by the Physics Dept. of Turin University (under the agreement with Consorzio Interuniversitario per la Fisica Spaziale) at the C3S Supercomputing Centre, Italy. We thank the staff of the GMRT that made these observations possible. GMRT is run by the National Centre for Radio Astrophysics of the Tata Institute of Fundamental Research. The National Radio Astronomy Observatory and Green Bank Observatory are facilities of the U.S. National Science Foundation operated under cooperative agreement by Associated Universities, Inc. This research made use of Matplotlib~\citep{hunter2007}, APLpy~\citep[an open-source plotting package for Python,][]{robitaille2012}, Astropy~\citep[a community-developed core Python package and an ecosystem of tools and resources for astronomy,][]{astropy2022}, SAOImageDS9~\citep[developed by Smithsonian Astrophysical Observatory,][]{ds9}.
\end{acknowledgements}

% WARNING
%-------------------------------------------------------------------
% Please note that we have included the references to the file aa.dem in
% order to compile it, but we ask you to:
%
% - use BibTeX with the regular commands:
%   \bibliographystyle{aa} % style aa.bst
%   \bibliography{Yourfile} % your references Yourfile.bib
%
% - join the .bib files when you upload your source files
%-------------------------------------------------------------------

\bibliographystyle{aa}
\bibliography{bibl}

\begin{appendix}
\label{appendix}

\section{Data calibration and imaging}
\label{appendix:data_calibration}
\subsection{LOFAR HBA data}
\label{sec:hba_data}
The calibration strategy used for the LOFAR-VLBI data at 144~MHz (observations listed in Tab.~\ref{tab:data}) is extensively described in Paper I. Given this, we refer the reader to Paper I for a detailed description of the Dutch stations calibration, and will focus our discussion here on the calibration of the international stations (IS) for the wide-field imaging. We followed the procedures described in~\citet{morabito2022} and~\citet{dejong2024}, with different ad-hoc adjustments, especially in the imaging step, required by the highly complex morphology of A2255, with the combination of extended and structured radio galaxies overlapping the cluster radio halo.

\begin{table*}[h!]
\centering
\caption{List of the observations used in this paper and final maps properties (Fig.~\ref{fig:lofar_widefield} and Fig.~\ref{fig:a2255_ugmrt_vlaconcat}).}
\label{tab:data}
\begin{tabular*}{\textwidth}{c @{\extracolsep{\fill}} ccccccc}
\hline\hline
Telescope & Obs. ID & Date & Band & On-source time & Robust & Restoring beam, PA & $\sigma_{\rm{rms}}$\\
 & & [YYYY-MM-DD] & [MHz] & & & & [$\rm{\muup Jy~beam^{-1}}$]\\
\hline \hline
\multirow{7}{*}{LOFAR} & L720378 & 2019-06-07 & \multirow{7}{*}{$120-168$} & 8 h & \multirow{7}{*}{$-1.5$} & \multirow{7}{*}{$1.5^{\prime\prime} \times 1.5^{\prime\prime}$, $0^{\circ}$} & \multirow{7}{*}{$38$}\\
 & L725454 & 2019-06-22 & & 8 h & & & \\
 & L726708 & 2019-06-28 & & 8 h & & & \\
 & L727110 & 2019-07-03 & & 8 h & & & \\
 & L733077 & 2019-08-09 & & 8 h & & & \\
 & L747613 & 2019-09-28 & & 8 h & & & \\
 & L751366 & 2019-10-04 & & 8 h & & & \\
\hline
\multirow{4}{*}{uGMRT} & 46\_063 & 2024-07-23 & \multirow{4}{*}{$1060-1460$} & $9.33$ h & \multirow{4}{*}{$-1.0$} & \multirow{4}{*}{$1.5^{\prime\prime} \times 1.5^{\prime\prime}$, $0^{\circ}$} & \multirow{4}{*}{$5$}\\
 & 48\_082 & 2025-05-29 &  & $7.2$ h &  &  & \\
 & 48\_082 & 2025-07-01 &  & $8.5$ h &  &  & \\
 & 48\_082 & 2025-09-14 &  & $8.5$ h &  &  & \\
\hline
VLA (C) & \multirow{2}{*}{14B-165} & 2014-10-26 & \multirow{3}{*}{$1008-2032$} & $4.46$~h & \multirow{3}{*}{$-0.5$} & \multirow{3}{*}{$1.3^{\prime\prime} \times 1.2^{\prime\prime}$, $94.5^{\circ}$} & \multirow{3}{*}{$3$}\\
VLA (B) & & 2015-02-28 & & $6.27$~h & & & \\
VLA (A) & 24B-067 & 2024-12-28 & & $8.65$~h & & & \\
\hline\hline
\end{tabular*}
\tablefoot{Column 1: telescope name; for the VLA, the array configuration is specified in brackets. Column 2: ID of the observation. Column 3: observation starting date. Column 4: observing band. Column 5: time on source. Column 6: robust parameter used for Briggs weighting during imaging. Column 7: restoring beam size and position angle (PA). Column 8: final rms noise.}
\end{table*}

\subsubsection{Direction-independent calibration of the ILT}
\label{sec:dical_ilt}
As already done in Paper I, solutions from the Dutch array corrected for both direction-independent (DIE) and direction-dependent effects (DDE) were transferred to the IS. Then, the datasets were concatenated into sub-bands of 1.95~MHz. Given that we aimed to image only the cluster field, and that the IS have a reduced FoV with respect to the Dutch stations because of the larger number of antenna tiles~\citep{morabito2022}, we subtracted all the sources outside a box region of $1.5^{\circ}\times 1.5^{\circ}$ using the sky model provided by the \textsc{ddf-pipeline}\footnote{\url{https://github.com/mhardcastle/ddf-pipeline}}~\citep{shimwell2019,tasse2021} at $6^{\prime\prime}$ resolution. The source subtraction was performed individually for each sub-band of each observation: to speed up this step, we split the subtraction among different computing nodes using the Pleiadi computing system~\citep[][]{taffoni2024}. Each of the available 24 nodes\footnote{36 CPUs Intel Xeon E5-2697 v4} is equipped with 256 GB of RAM and 8 TB of local storage. This allowed us to subtract multiple sub-bands in parallel, gaining a factor four in wall-clock time for the subtraction of a single observation. The resulting subtracted datasets were then used for further calibration of the IS.
\par We selected 4C $+64.21$~\citep[RA: $\rm{17^h 19^m 59^s}$, Dec: $\rm{64^{\circ}04'37^{\prime\prime}}$,][]{pilkington1965} as the most suitable delay calibrator from the Long Baseline Calibrator Survey~\citep[LBCS,][]{moldon2015,jackson2016,jackson2022}, a powerful radio source, compact on arcsec-scale, with flux density of $\sim 4.87~\rm{Jy}$ at 144~MHz. In Paper I, we detailed the initial sky model that we built and used for self-calibration on this source. We used \textsc{facetselfcal}~\citep{vanweeren2021}, which combines the Default Preprocessing Pipeline~\citep[\textsc{DP3},][]{vandiepen2018,dijkema2023} and \textsc{WSClean}~\citep{2014offringa,offringa2017} to perform self-calibration. The adopted strategy follows the one described in~\citet{dejong2024}, with some adjustments made for the characteristics of our calibrator.
\begin{enumerate}
\item \texttt{scalarphasediff}: solution interval 8 min, frequency smoothness 10~MHz;
\item \texttt{scalarphase}: 32 s, frequency smoothness increasing at each step from 2 to 20~MHz;
\item \texttt{complexgain}: 20 min, frequency smoothness 7.5~MHz;
\item \texttt{scalaramplitude}: 120 min, over the whole band.
\end{enumerate}
We ignored all baselines shorter than $40~\rm{k}\lambda$, corresponding to an angular scale of about $5^{\prime\prime}$ at 144~MHz, using the parameter \texttt{uvmin}, to prevent possible incompleteness in the sky model. Through \textsc{facetselfcal}, we also averaged the input data down to 32~s in time and 488~kHz in frequency and used a Briggs weighting with robustness of $-1.5$ for the imaging~\citep{1995briggs}. At the end of the self-calibration routine on the in-field calibrator, one Hierarchical Data Format 5 solution file (\texttt{h5parm}) for each calibration step and one containing all the solutions merged, with phases and amplitudes corrections, are produced. The best solutions for visibilities were obtained after 19 self-calibration cycles.

\subsubsection{Direction-dependent calibration of the ILT}
\label{sec:decal_ilt}
After the DIE correction for the IS there are still residual DDE across the FoV, mainly due to the ionosphere and errors in the beam model, that need to be corrected. The standard strategy to correct for DDEs in the LOFAR-VLBI wide-field imaging is to select a number of direction-dependent calibrators across the field, dividing it into as many smaller facets, calibrate against these sources, and apply the resulting solutions across the entire resident facet~\citep[the so-called \enquote{faceting} technique,][]{2016vanweeren}. To choose suitable DDE calibrators, we ran \textsc{PyBDSF}\footnote{\url{https://pybdsf.readthedocs.io/en/latest/}}~\citep{mohan2015} on a preliminary wide-field, DIE corrected, image at $1.5^{\prime\prime}$ resolution, obtaining a catalog of the sources in the field. Among all the detected sources, we selected only those with a peak surface brightness above $20~\rm{mJy~beam^{-1}}$ within the selected area. A further visual selection was made based on the source morphology and flux (more compact and brighter), to remove bad candidate DDE calibrators beforehand.
\par
\begin{figure}[h!]
\centering
\includegraphics[width=\columnwidth]{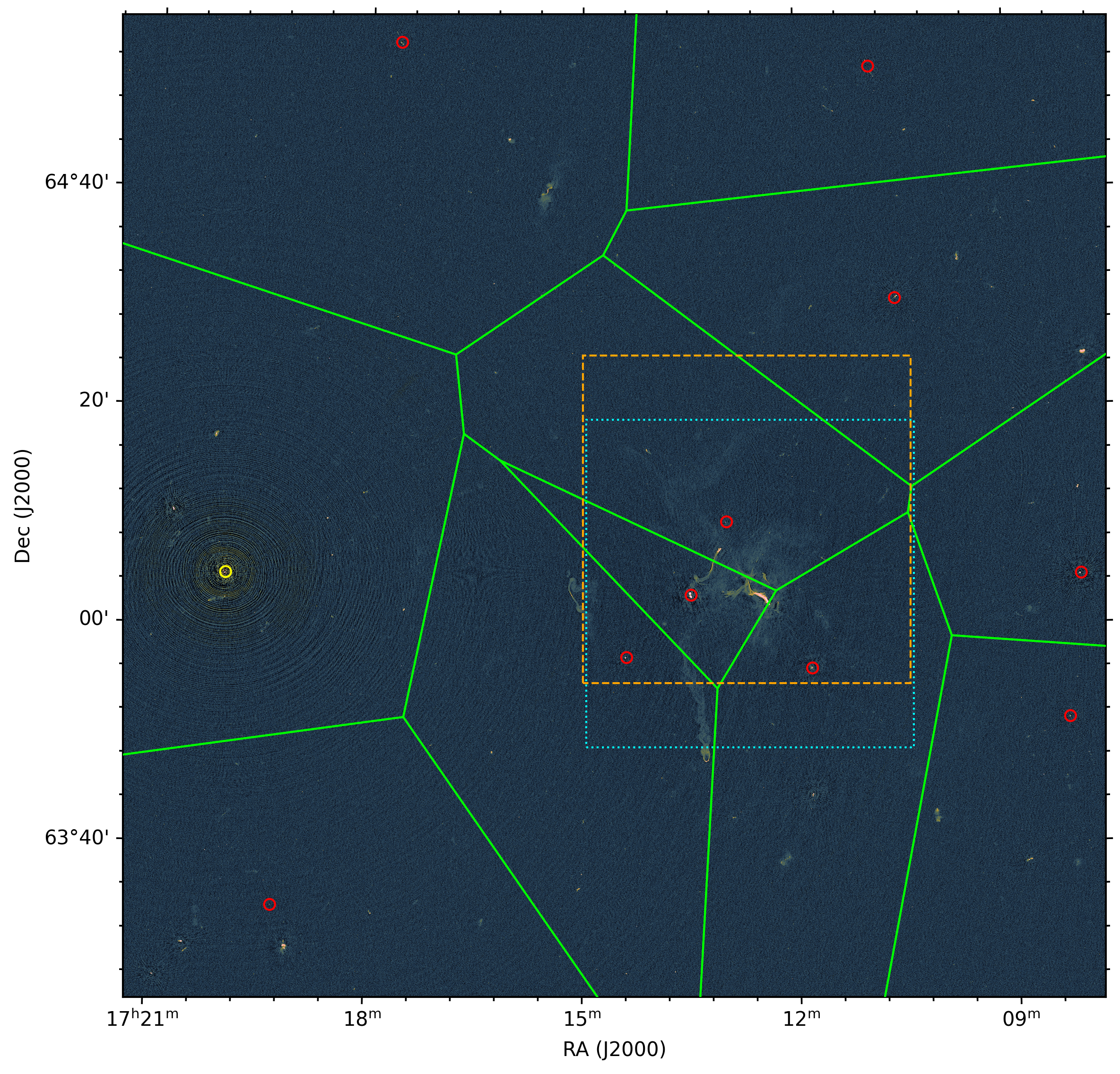}
\caption{The facets layout (in green) for wide-field imaging with LOFAR-VLBI. In red we highlighted the position of the direction-dependent calibrators, while in yellow the delay-calibrator. We also indicated, with squares, the position and size of the full uGMRT (in cyan, dotted line) and VLA (in orange, dashed line) maps.}
\label{fig:facets}
\end{figure}

 We detail hereafter the calibration strategy to deal with DDE: all these steps have been done for each observation individually. We phase-shifted the subtracted datasets towards the direction of the candidate DDE calibrator for each sub-band, averaging the datasets down to 195.31~kHz and 16 seconds. Then, solutions from DIE calibration of the delay calibrator were applied to all the sub-bands, which were finally concatenated to obtain a unique dataset for each calibrator. Every DDE calibrator candidate was self-calibrated with \textsc{facetselfcal}, firstly using the automated settings and then, where necessary, refining the calibration with a better suited strategy. With the \texttt{auto} mode, the script is optimized to set automatically several calibration parameters, such as the solution interval. In case of divergent or not improving solutions with increasing self-calibration cycles we manually tweaked the solution interval and the frequency smoothness. We used as solution types two rounds of \texttt{scalarphase}, with the first one that resets the solution to $0.0$ (for the phases) and $1.0$ (for the amplitudes) for the Dutch stations~\citep{dejong2025calibration}, with an increasing solution interval ranging from 1-5 minutes. Then, we also used a \texttt{scalarcomplexgain} solve with a longer solution interval (120 minutes) and larger frequency smoothness (10-20~MHz). After a careful evaluation of self-calibration solutions from inspection plots and from final images, we ended up with 10 direction-dependent calibrators that were used to correct for DDE (listed in Tab.~\ref{tab:ddcals} and encircled in red in Fig.~\ref{fig:facets}).
\par One of the selected direction-dependent calibrators (J171259+640931) is located, in projection, just above the cluster center. After applying the solutions from this calibrator to its resident facet, the radio galaxies that populate the cluster center still showed several artifacts, requiring additional calibration refinement. We used the Double radio galaxy as an additional calibrator (J171329+640249), adding a secondary layer of calibration to adjust the phases for the central region of the cluster. On the Double, we used two rounds of \texttt{scalarphase} with solution interval of 5 minutes. The first one is in charge of resetting the solutions for the Dutch stations, while the second uses a smoothness in frequency over the whole bandwidth. Moreover, given the extended morphology of the source, we enabled the \texttt{multiscale} deconvolution algorithm~\citep{offringa2017}, setting a maximum number of seven, automatically selected, scales for the cleaning. The output solutions resulted in improved solutions quality for the most crowded region of the field.

\begin{table}
\centering
\caption{List of selected direction-dependent calibrators.}
\label{tab:ddcals}
\begin{tabular}{ccc}
\hline\hline
RA (J2000) & Dec (J2000) & $\langle\sigma_{\rm_{rms}}\rangle$\\
 & & [$\rm{\muup Jy~beam^{-1}}$]\\
\hline
$\rm{17^{h}13^{m}29^{s}}$ & $64^{\circ}02'49^{\prime\prime}$ & $35$\\
$\rm{17^{h}08^{m}02^{s}}$ & $64^{\circ}04'24^{\prime\prime}$ & $41$\\
$\rm{17^{h}08^{m}14^{s}}$ & $63^{\circ}51'17^{\prime\prime}$ & $41$\\
$\rm{17^{h}10^{m}35^{s}}$ & $64^{\circ}29'53^{\prime\prime}$ & $35$\\
$\rm{17^{h}10^{m}55^{s}}$ & $64^{\circ}51'07^{\prime\prime}$ & $42$\\
$\rm{17^{h}11^{m}48^{s}}$ & $63^{\circ}56'04^{\prime\prime}$ & $36$\\
$\rm{17^{h}12^{m}59^{s}}$ & $64^{\circ}09'31^{\prime\prime}$ & $37$\\
$\rm{17^{h}14^{m}23^{s}}$ & $63^{\circ}57'07^{\prime\prime}$ & $33$\\
$\rm{17^{h}17^{m}36^{s}}$ & $64^{\circ}53'21^{\prime\prime}$ & $44$\\
$\rm{17^{h}19^{m}17^{s}}$ & $63^{\circ}34'13^{\prime\prime}$ & $42$\\
\hline\hline
\end{tabular}
\tablefoot{Column 1: right ascension (in J2000). Column 2: declination (in J2000). Column 3: average rms noise of the final image from self-calibration.}
\end{table}

\begin{figure*}[h!]
\centering
\includegraphics[width=\textwidth]{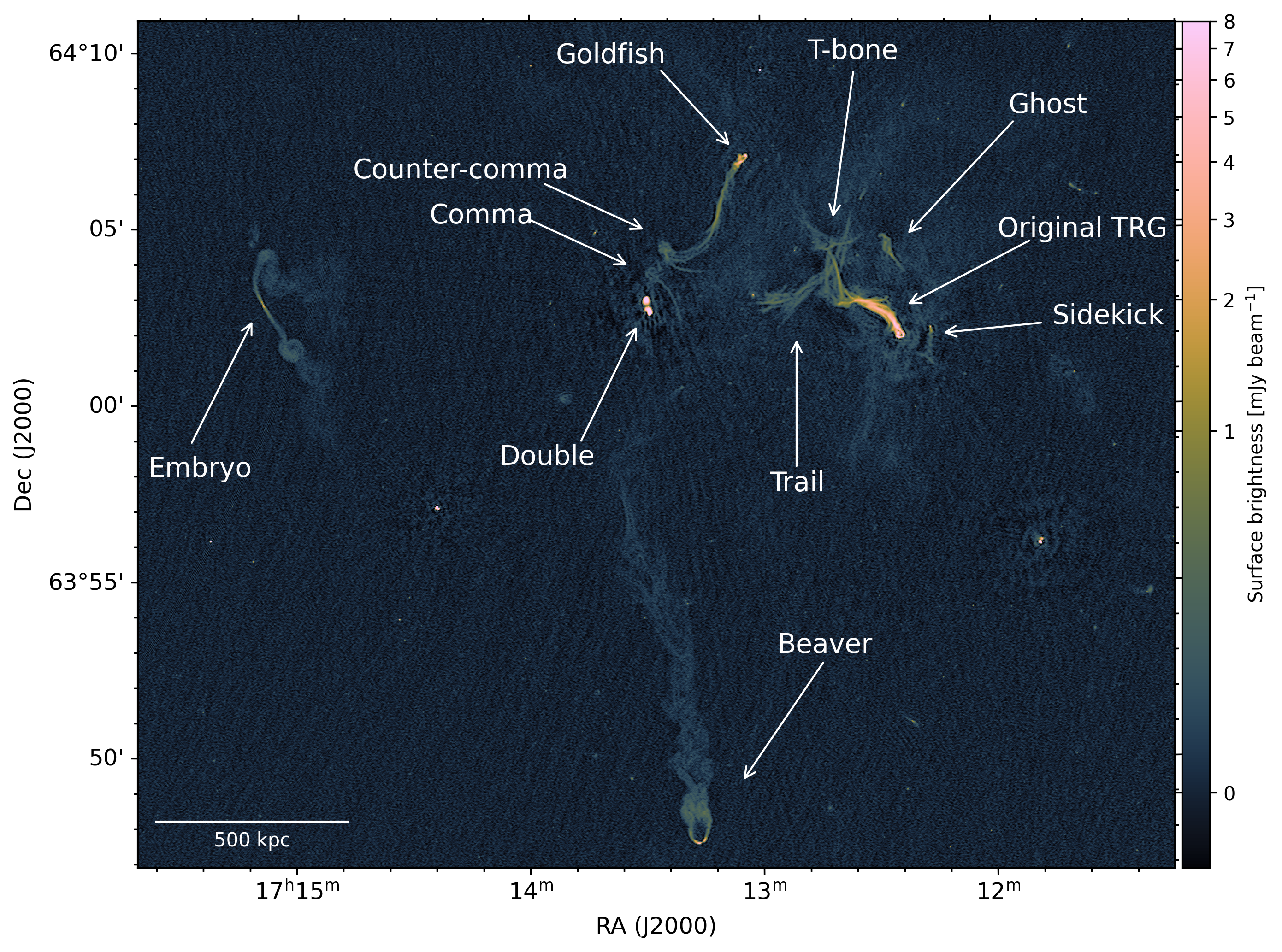}
\caption{LOFAR-VLBI wide-field image of A2255 at 144~MHz. This image of the cluster comprehends the major cluster member radio galaxies and diffuse features, labeled in white following the nomenclature in~\citet{botteon2020}. The final image has a resolution of $1.5^{\prime\prime}$ resolution, with rms noise of $38~\rm{\muup Jy~beam^{-1}}$.}
\label{fig:lofar_widefield}
\end{figure*}

\subsubsection{Wide-field imaging}
\label{sec:widefield_imaging}
The wide-field image was produced with an angular resolution of $1.5^{\prime\prime}$, covering a field of view of $1.5^{\circ}\times1.5^{\circ}$, centered on RA $17^{\rm{h}}14^{\rm{m}}33^{\rm{s}}.03$ and Dec $+64^{\circ}11'04^{\prime\prime}.10$. The different observations, which have been calibrated individually, are finally combined in this step. Before imaging, we averaged the input datasets by a factor of four in both time and frequency, which means a final time and frequency resolution of 4~s and 48.84~kHz, respectively. This ensures to have negligible smearing effects at the imaging resolution~\citep{dejong2024}, especially considering that most of the target sources of our analysis are close to the pointing center. To apply the corrections found for DDE calibration, we used the facet-based imaging in \textsc{WSClean}, enabled by the parameter \texttt{facet-regions}, which requires a DS9~\citep{ds9} region file that contains all the facets used for wide-field imaging, produced through a Voronoi tessellation, wherein the calibration solutions are assumed to be constant~\citep{schwab1984}. The layout of the facets, together with the position of the resident direction-dependent calibrators, is shown in Fig.~\ref{fig:facets}. We also merged, for each observation, the h5parm solution files coming from all the selected DDE calibrators using the \textsc{h5\_merger} script~\citep{dejong2022}: the h5parm merged files, for each observation, are used in \textsc{WSClean} through the \texttt{apply-facet-solutions} option.
\par The main difficulty in producing such images for a cluster field, especially for A2255, is the coexistence of diffuse emission up to several Mpc scales, which becomes more relevant with increasing sensitivity and enhances the local rms noise, and individual radio galaxies, especially with their thin ($\approx 4-10$ kpc) filamentary tails. If the rms noise is not correctly evaluated locally, the CLEANing~\citep{1974hogbom} process might not be deep enough for the sources embedded in the radio halo. Moreover, the tails and filaments are naturally disadvantaged during CLEAN because of their shape, which is elongated only along one direction. All of these points required specific settings and several imaging attempts to fine-tune the imaging strategy to be used in \textsc{WSClean}.
\par For the final image (size $13,500 \times 13,500$ pixels, with $0.4^{\prime\prime}$ pixel scale), we used a spatially varying rms noise, evaluated within regions of 200 point spread functions (PSF) each, and enabled direction-dependent PSFs with a size of 3 cells in each direction; we employed a Briggs weighting scheme with robust $-1.5$, automatic masking, multiscale deconvolution and parallel reordering and deconvolution. We imaged the whole FoV view at once, since the image fits the memory of a single computing node. After deconvolution, we corrected the image flux density scale using the in-field calibrator as a reference, given that it is a well known radio source on multiple frequencies: comparing the flux density at 144~MHz between the $6^{\prime\prime}$ and the $1.5^{\prime\prime}$ image, we found a correction factor of 1.067, which we applied to the whole image. The use of a sky model for the in-field calibrator already provided reasonably good astrometric accuracy~\citep[within $0.06^{\prime\prime}$ over the entire FoV with respect to the positions of the extragalactic objects provided by Gaia Data Release 3,][]{gaia2023}. The final image, shown in Fig.~\ref{fig:lofar_widefield} (with a zoom on the central region in Fig.~\ref{fig:lofar_widefield_zoom}), has a rms noise of $38~\rm{\muup Jy~beam^{-1}}$, with the main features labeled following the nomenclature in~\citet{botteon2020}. This represents the deepest and only LOFAR-VLBI wide-field image ever obtained targeting a galaxy cluster, and will be used for high-resolution spectral studies later in this paper.

\begin{figure*}[h!]
\centering
\includegraphics[width=\hsize]{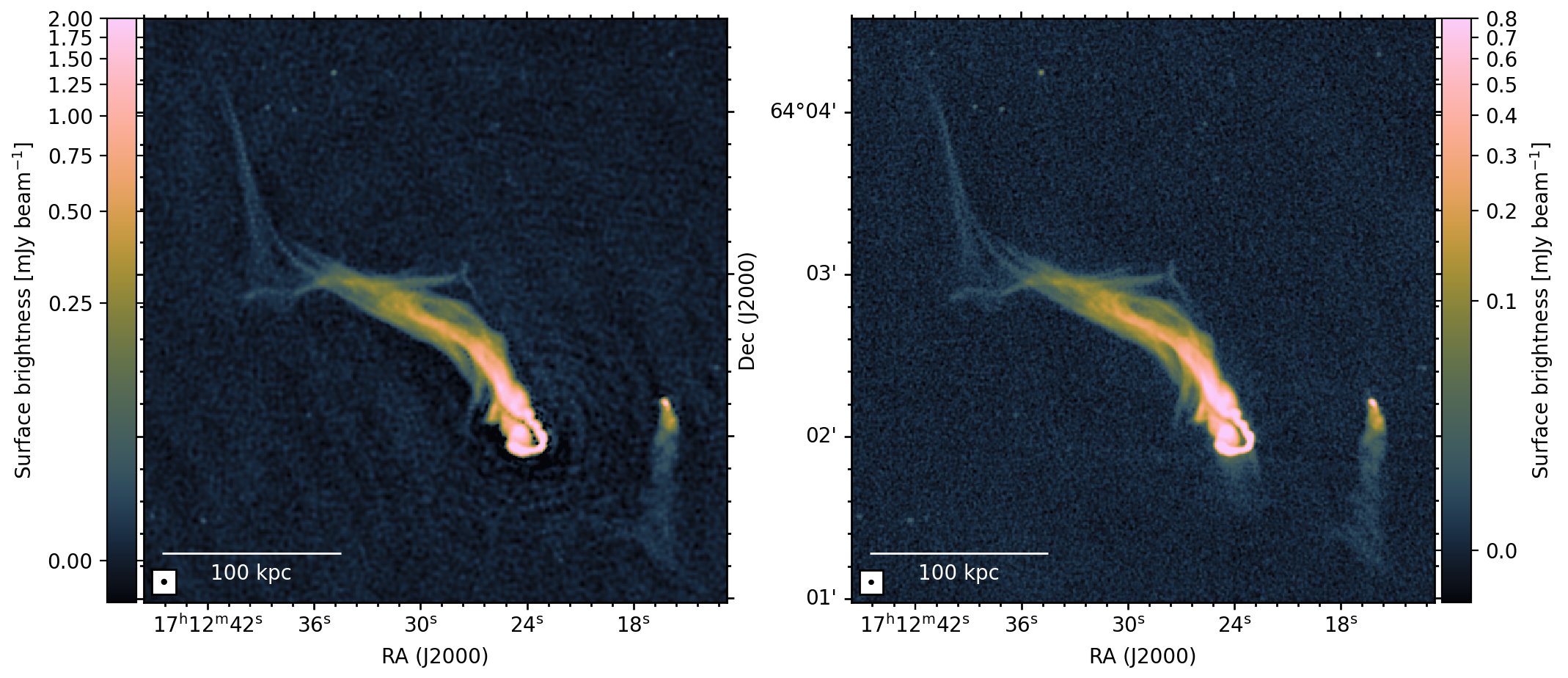}
\caption{Original TRG radio maps at GHz frequencies. Left: uGMRT map of A2255. Total intensity image in Band 5 (central frequency 1260~MHz), resolution $1.5^{\prime\prime} \times 1.5^{\prime\prime}$, rms noise $5~\rm{\muup Jy~beam^{-1}}$. Right: VLA map of A2255. Total intensity image in L-band (central frequency 1520~MHz), resolution $1.3^{\prime\prime} \times 1.2^{\prime\prime}$, rms noise $3~\rm{\muup Jy~beam^{-1}}$.}
\label{fig:a2255_ugmrt_vlaconcat}
\end{figure*}

\subsection{uGMRT Band 5 data}
\label{sec:ugmrt}
We observed A2255 using the uGMRT in Band 5 ($1060-1460$~MHz) for around 33 hours on target (Tab.~\ref{tab:data}, proposal codes $46\_063$ and $48\_082$, P.I. De Rubeis). The bandwidth is divided into 4096 channels, each with a resolution of 97.656~kHz. For the first observation, both the uGMRT Wideband Backend (GWB) and the GMRT software Backend (GSB) were available; for the others, we used only the GWB. We processed all the data using the Source Peeling and Atmospheric Modeling pipeline~\citep[\textsc{SPAM},][]{intema2009}. For each observation, the wideband data were split into six sub-bands, spanning around $66$~MHz each, which were independently processed. After flux density scale calibration and bandpass (primary calibrators: 3C48 for the first observation, 3C286 for the second observation), using the flux density scale model from~\citet{scaife2012}, the pipeline corrects for direction-independent gains using a starting global sky model obtained from GSB data. Given the lack of a sufficient number of sources to peel for the direction-dependent self-calibration, we settled with the direction-independent ionospheric corrections. To produce deep, wideband images, we combined the calibrated sub-bands and imaged the data with \textsc{WSClean}. We used the \texttt{multiscale} deconvolution algorithm, and automatic masking for deconvolution with a spatially varying rms noise for masking. The final image, whose zoom around the Original TRG is shown in the left panel of Fig.~\ref{fig:a2255_ugmrt_vlaconcat}, central frequency 1.26~GHz, was obtained using a Briggs \texttt{robust} of $-0.5$ to reach a noise level of $5~\rm{\muup Jy~beam^{-1}}$.

\subsection{VLA L-band data}
\label{sec:vla}
A2255 was observed with the VLA in L-band ($1008-2032$~MHz) in A-configuration for a total amount of 10 hours (Tab.~\ref{tab:data}, proposal code 24B-067, P.I. De Rubeis). The bandwidth is divided into 1024 channels, each with a resolution of 1000~kHz. We processed the data using the \textsc{VLA Calibration Pipeline v2024.1.0.8}, which uses the Common Astronomy Software Applications~\citep[\textsc{CASA v6.6.5.31}][]{casa2022} to perform total intensity standard flagging and calibration procedures. We used 3C286 as the primary calibrator for bandpass, using the flux density scale model from~\citet{perley2017}, and J1634+627 as the secondary calibrator for gains. Subsequent to reviewing the calibration outcomes from the National Radio Astronomy Observatory's (NRAO) automatic data processing, it became necessary to re-execute the pipeline for supplementary data flagging. The primary calibrator was also used for the calibration of the absolute polarization angle: we modeled the polarization angle as in~\citet{hugo2024}, which observed a drop in the intrinsic angle of 3C286 approaching the lower bound of the L-band. The secondary calibrator was used instead to calibrate the on-axis instrumental leakage. We made a polynomial fit to its known values of linear polarization fraction and polarization angle to build a frequency-dependent polarization model, following the NRAO polarimetry guide~\footnote{\url{https://science.nrao.edu/facilities/vla/docs/manuals/obsguide/modes/pol}}. The final calibration tables were applied to the target. Radio frequency interference (RFI) was removed both manually and using statistical flagging \textsc{CASA} algorithms also from the cross-correlation products. For Stokes I imaging, together with data in A-configuration, we added archival data of the same target in B- and C-configuration (proposal code 14B-165, P.I. L. Rudnick), to add short spacings and increase the sensitivity, especially to the more extended emission of the filaments. The calibration strategy for these data will be detailed in an upcoming publication (Rajpurohit et al. \textit{in preparation}). We concatenated the measurement sets and re-calculated the weights based on the new data. Then we imaged the concatenated visibilities using \textsc{WSClean 3.5}. The final image was obtained using a Briggs weighting scheme, with \texttt{robust=-0.5}, and \texttt{multiscale} deconvolution algorithm. A zoom in on the Original TRG is shown in the right panel of Fig.~\ref{fig:a2255_ugmrt_vlaconcat}. It has an rms noise of $\sim 3~\rm{\muup Jy~beam^{-1}}$, a final synthesized beam of $1.3^{\prime\prime} \times 1.2^{\prime\prime}$. We also performed several rounds of self-calibration to further improve the final image, but since these efforts did not yield significant improvements, the self-calibration results were not included in the paper.

\section{Flux density scale}
\label{appendix:flux_density_scale}
For the spectral index analysis, we checked for any possible flux density scale offset in the uGMRT observations using the VLA as a reference. This choice is motivated by the larger amount of works in the literature in which the VLA L-band was used, with respect to the less tested uGMRT Band 5. We produced maps with the same imaging parameters and inner $uv$-cut for both arrays, and obtained catalogs using \textsc{PyBDSF}. We cross-matched the two catalogs, using $1^{\prime\prime}$ as maximum separation. We then calculated the spectral index between 1260 and 1520~MHz and selected only the sources having a spectral index in the range $-0.5<\alpha<1.5$, to take into account also sources with inverted spectrum. We also set a limit in the uGMRT signal-to-noise ratio to be over $5$ and further reduced our sample considering only compact sources identified by a single-Gaussian component. This selection was made to isolate all the compact sources observed with both uGMRT and VLA, which display a spectral index distribution with median $\rm{med(\alpha)} = 0.578$. Considering, as stated before, the VLA flux density scale as a reference, we re-scaled the uGMRT flux density scale with an upward scaling factor ($8\%$) which returns a median of the spectral indices of $0.75$, usually found in large samples of radio sources~\citep[e.g.,][]{mahony2016}.

\section{Spectral index error map}
\label{appendix:spidx_error}
Spectral index error maps related to Fig.~\ref{fig:spidx}.
\begin{figure}[h!]
\centering
\includegraphics[width=\columnwidth]{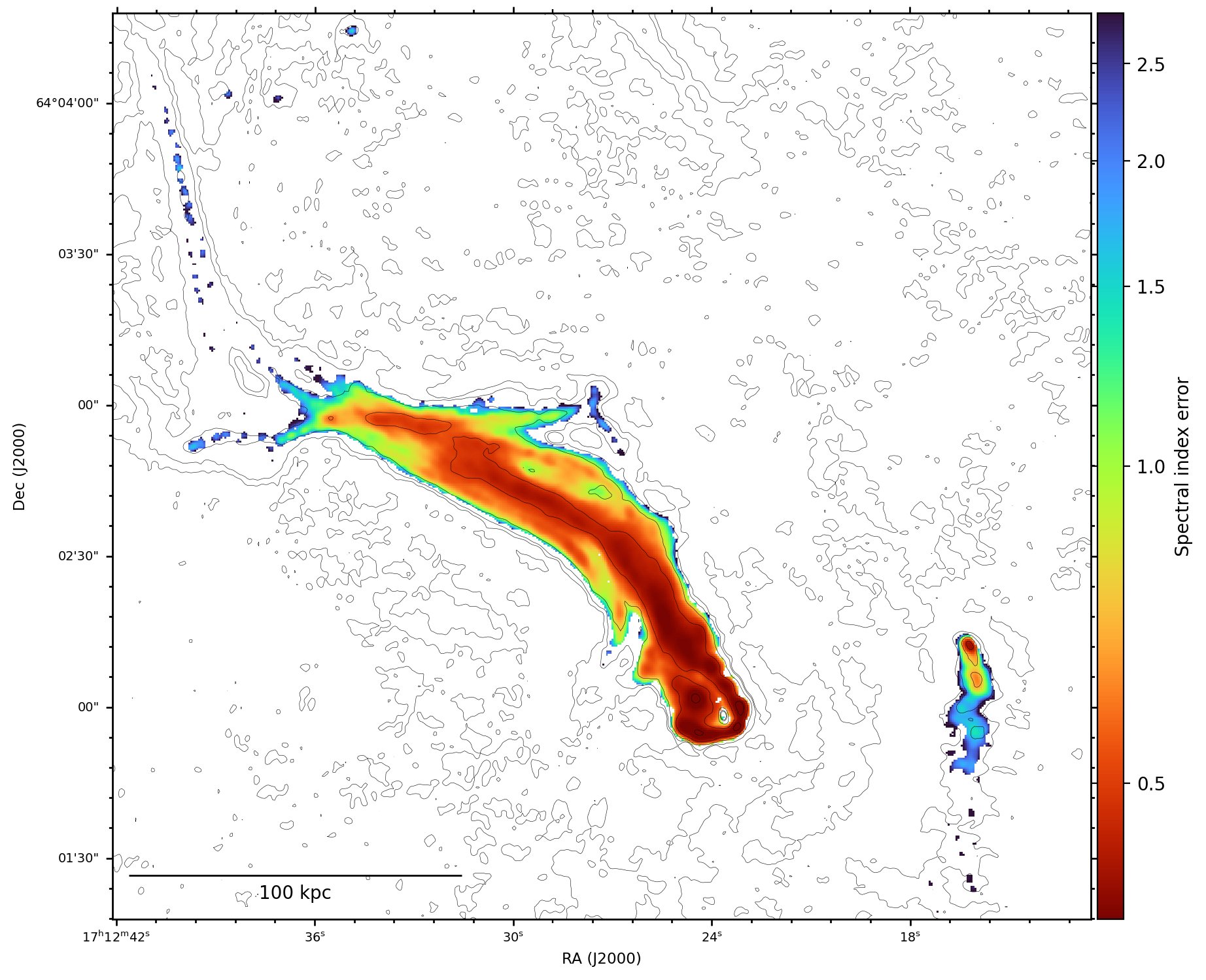}
\caption{Spectral index error map, referred to Fig.~\ref{fig:spidx}, obtained using LOFAR-VLBI (144~MHz), uGMRT (1260~MHz), and VLA (1520~MHz) data for the Original TRG. Grey contours represent $3,~6,~12,~24\ldots\sigma_{\rm{rms}}$ levels, with $\sigma_{\rm{rms}} = 38~\rm{\muup Jy~beam^{-1}}$ being the noise from the LOFAR-VLBI, wide-field image at $1.5^{\prime\prime}$ resolution in Fig.~\ref{fig:lofar_widefield_zoom}.}
\label{fig:spidx_err}
\end{figure}

\newpage

\section{Regions for surface brightness and spectral index trends}
\label{appendix:regions}
\begin{figure}[h!]
\centering
\includegraphics[width=\columnwidth]{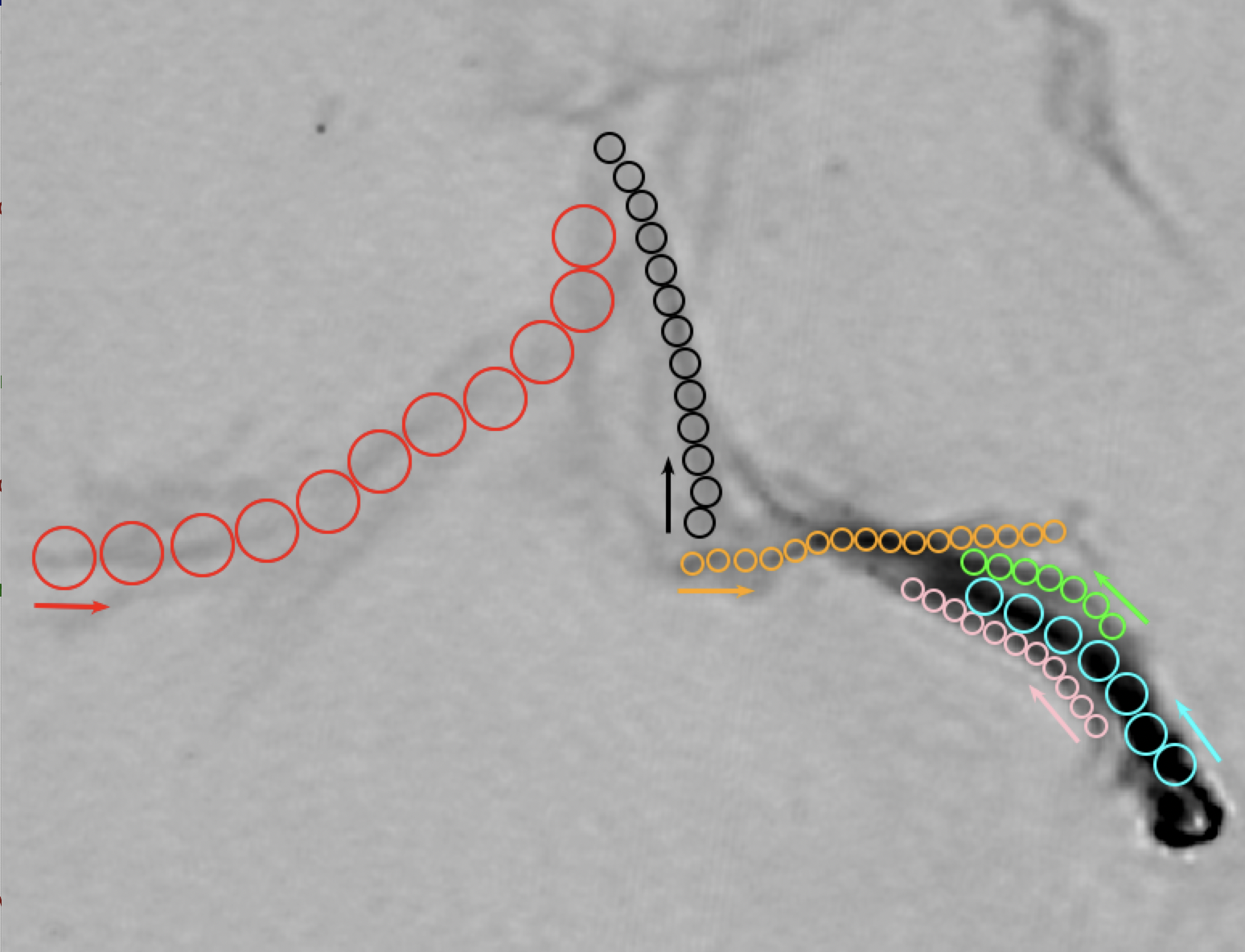}
\caption{Regions used to evaluate the average surface brightness and spectral index in Fig.~\ref{fig:trends}, with different colors corresponding to different features. These are the main tail (cyan), the horizontal (orange) and vertical (black) filaments, F1 (pink), F2 (green), and the Trail (red). The arrows indicate the direction of the trends. A grayscale version of the map from Fig.~\ref{fig:lofar_widefield} is displayed.}
\label{fig:trends_regions}
\end{figure}

\section{Rotation measure values distribution}
\label{appendix:rm_distribution}
\begin{figure}[h!]
\centering
\includegraphics[width=\columnwidth]{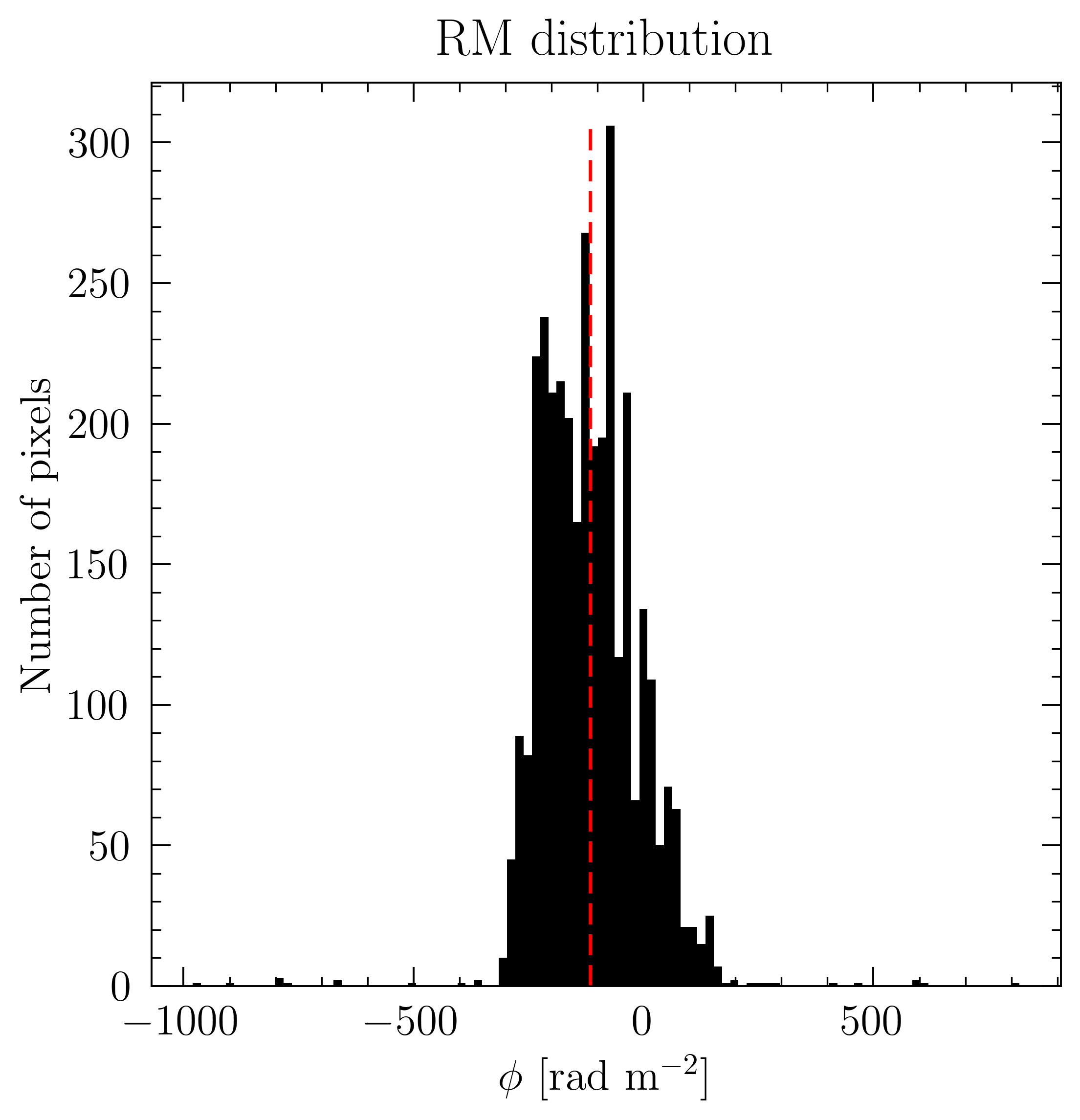}
\caption{RM distribution for the Original TRG, considering only pixels above $5\sigma_{\rm{P}}$ (from Fig.~\ref{fig:polarization}). The vertical dashed red line indicates the average $\langle\rm{RM}\rangle = -115.7~\rm{rad~m^{-2}}$ resulting from Gaussian fitting, while the standard deviation is $\sigma_{\rm{RM}} = 107.4~\rm{rad~m^{-2}}$.}
\label{fig:rmdistribution}
\end{figure}
\end{appendix}

\end{document}